\documentclass[12pt]{article}

\usepackage{amsmath,amsthm,amssymb,euscript,epsf,epsfig,cite}
\usepackage{array,longtable}
\usepackage{fancybox}
\usepackage{comment} 

\usepackage{rotating}

\usepackage{tikz,pbox}
\usetikzlibrary{shapes,arrows}
\usetikzlibrary{calc,decorations.markings}

\usepackage{hyperref}

\makeatletter
\@addtoreset{equation}{section}
\makeatother

\def\one{{\hbox{ 1\kern-.8mm l}}}
\def\zero{{\hbox{ 0\kern-1.5mm 0}}}

\def\mC{ \mathbb{C}}
\def\Div{ { \rm { Div } } } 
\def\bq{ { \bf {q} } }
\def\br{ { \bf{r} } }

\def\Aut{ \hbox{Aut}}

\def\s{ \sigma}

  \def\cC{{\cal C}}
  
 \def\cH{{\cal H}} 
  
 \def\cN{{\cal N}} \def\cO{{\cal O}}
\def\cP{{\cal P}}

 \def\cZ{{\cal Z}}

\def\Sym{ \hbox{Sym} } 
 
\def\Aut{ {\rm Aut} }

\newcommand{\be}{\begin{equation}}
\newcommand{\ee}{\end{equation}}
\newcommand{\beq}{\begin{equation}}
\newcommand{\eeq}{\end{equation}}
\newcommand{\bea}{\begin{eqnarray}\displaystyle}
\newcommand{\eea}{\end{eqnarray}}

\def\s{ \sigma }

\def\Diag{ \rm{Diag} }

\newcommand{\Tr}{{\rm Tr}}
\newcommand{\tr}{{\rm tr}}

\begin{document}

\begin{flushright}
QMUL-PH-21-21,  LIMS-2021-008
\end{flushright}

\medskip

\begin{center}

{\Large \bf Integrality, Duality and Finiteness in}
 
{\Large \bf   Combinatoric topological strings} 

\bigskip

Robert de Mello Koch$^{a,b\dag}$, Yang-Hui He$^{c, \dag \dag}$, Garreth Kemp$^{d,*}$,  Sanjaye Ramgoolam$^{e,b,**}$ 

\bigskip

{\small
$^{a}${\em Guangdong Provincial Key Laboratory of Nuclear Science$^1$, Institute of Quantum Matter, 
South China Normal University, Guangzhou 510006, China}\\
$^a${\em Guangdong-Hong Kong Joint Laboratory of Quantum Matter$^1$, Southern Nuclear Science Computing centre, 
South China Normal University, Guangzhou 510006, China}\\
$^{b}${\em  School of Physics and Mandelstam Institute for Theoretical Physics,}
{\em University of Witwatersrand, Wits, 2050, South Africa} \\
\medskip
$^{c}$ {\em  London Institute for Mathematical Sciences, Royal Institution of GB, W1S 4BS, UK\\
$^{c}$ Merton College, University of Oxford, OX14JD, UK \\
$^{c}$ Department of Mathematics, City, University of London, EC1V 0HB, UK\\
$^{c}$ School of Physics, NanKai University, Tianjin, 300071, P.R. China} \\
\medskip
$^{d}${\em Department of Mathematics and Applied Mathematics,}\\
{\em University of Johannesburg, Auckland Park, 2006, South Africa}\\
\medskip
$^{e}$
{\em School of Physics and Astronomy,  Centre for Research in String Theory}\\
{\em Queen Mary University of London, London E1 4NS, United Kingdom }\\
}

\begin{abstract}

A remarkable result at the intersection of number theory and group theory states that the order of a finite  group $G$ (denoted $|G|$)  is divisible by the dimension  $d_R$ of any irreducible complex representation  of $G$.  
We show that the integer ratios
 ${ |G|^2 / d_R^2 } $ are combinatorially constructible using  finite  algorithms which take as input the amplitudes of  combinatoric topological strings ($G$-CTST) of finite groups based on 2D  Dijkgraaf-Witten topological field  theories ($G$-TQFT2).  
The ratios are also shown to be eigenvalues of handle creation operators in $G$-TQFT2/$G$-CTST. 
These strings have recently been discussed as toy models of wormholes and baby universes by Marolf and Maxfield, and Gardiner and Megas.   Boundary amplitudes of the  $G$-TQFT2/$G$-CTST provide algorithms for  combinatoric constructions of normalized characters. 
Stringy S-duality for closed $G$-CTST gives a dual expansion generated by disconnected entangled surfaces. 
There are  universal  relations between $G$-TQFT2 amplitudes due to the finiteness of the number $K $ of conjugacy classes. These relations  can be  labelled by Young diagrams and are captured by  null states in an inner product constructed by coupling the $G$-TQFT2 to a universal TQFT2 based on symmetric group algebras.  We discuss the scenario of a 3D holographic dual for this coupled theory and the implications  of the scenario for the factorization puzzle of 2D/3D holography raised by wormholes in 3D.  

\end{abstract}

\end{center}

\vskip.2cm 

\noindent 

{\small{ 
\noindent 
 { E-mails:  $^{\dag}$robert@neo.phys.wits.ac.za,${}^{\dag \dag}$hey@maths.ox.ac.uk, ${ }^{*}$garry@kemp.za.org,\\
 ${}^{**}$s.ramgoolam@qmul.ac.uk } 
} }

\newpage 

\tableofcontents

\section{Introduction}

A well-known fact in finite group theory equates  the sum of squares of dimensions of irreducible representations (irreps) to the order of the group $G$. Letting $R$ label the irreps and denoting the dimensions by $d_R$, we have 
\bea
|G| = \sum_R d_R^2 
\eea
where $ |G|$ is the order of  the group. Another well-known fact equates  the number of irreps to the number of conjugacy classes. These are two properties of  the set of irreps  which can be constructed using the combinatorics of  group elements and their group multiplication. It is natural to ask whether this combinatoric  constructibility of 
properties of $d_R $ goes further to allow reconstruction of all the individual $d_R$.  A remarkable fact about finite groups is that $ {|G | \over d_R } $ is an integer for every $R$. The proof, which relies on properties of algebraic integers, involves the intersection of number theory and group theory (see for example \cite{Simon} for the proof). In this paper, we will describe an algorithm for the construction of the integers $ |G|^2/d_R^2$, and hence of $ d_R$, from the combinatoric data of group multiplications. The multiplications are  shaped according to the fundamental groups of two dimensional surfaces.

Many interesting results in representation theory have combinatoric constructions. For example the enumeration of
representations of the symmetric group can be done by enumerating Young diagrams. The computation of dimensions $d_R$ for irreps of symmetric groups can be done by counting standard Young tableaux. The Littlewood-Richardson coefficient can be computed by a combinatoric rule for composing Young diagrams. These results are described in standard textbooks on representation theory, e.g. \cite{FultonHarris}. Further results along these lines are given in \cite{BarceloRam}. A number of open problems in representation theory revolve around finding combinatoric interpretations for representation theoretic quantities \cite{Stanley}. Such interpretations have implications for computational complexity theory 
\cite{MulVar,burg,BCI2011,IkMuWa-VanKron,PakPanova}.  In a recent paper \cite{QMRibb}, it was shown  that stringy combinatoric structures, notably bipartite ribbon graphs, 
can be used to provide a lattice interpretation for Kronecker coefficients.  The stringy nature of bipartite graphs reveals itself in a number of ways \cite{dMKR,JRR,HHJPRR,HeRe,HeHiPe}.

Here we turn to the question of whether string theory can provide an avenue for the constructibility of $ d_R$ and  $ {|G| \over d_R } $.  A number of developments in topological field theory and topological string theory,  provide valuable hints in this direction. Our constructions will be based on the topological field theory of flat $G$-bundles on two dimensional surfaces, which has concrete realizations as lattice constructions \cite{DW,Witten2dYM,FQ,FHK}. We will refer to this topological field theory based on $G$ as $G$-TQFT2. Recent work in connection with wormhole physics and baby universes  \cite{MarMax,GardMeg} has introduced sums over surfaces weighted by a string coupling $g_{st}$, where each surface supports a $G$-TQFT2, thus defining topological string theories based on $G$-TQFT2. We will refer to these string theories as combinatoric topological string theories or $G$-CTST. We will give a construction of $ \left ( { |G|/d_R } \right )^2$
which involves collecting  $G$-TQFT2 data from surfaces surfaces of different genera, so the construction  may be  naturally interpreted in terms of  $G$-CTST.  
 $G$-TQFT2 have been  used as an alternative approach to proving the integrality of the  ratios  $|G|/d_R$ in the mathematics literature in \cite{KWG}.

 While the paper starts with these motivations based on representation theoretic construction algorithms related to $G$-TQFT2/$G$-CTST, we then turn to physical questions related to these theories and it turns out that these ratios continue to play a key role.  We describe   transition probabilities constructed from $G$-TQFT2, from disjoint unions  of circles to disjoint unions of circles,   using the topological lattice formulation as a model for a two-dimensional path integral.  As usual the probabilities are squares of amplitudes computed from the path integral. The algebraic  structure of $G$-TQFT2 ensures that the amplitudes themselves are sums  over sectors labelled by irreducible representations of $G$. The weights include  the Plancherel distribution over irreps of finite groups \cite{BorOk} and various generalizations (depending on the choice of genus of the interpolating surfaces), which find a geometrical interpretation as amplitudes in $G$-TQFT2. The sums over sectors labelled by irreps are interpreted following \cite{MarMax} as sums over the $\alpha$-states, which were identified  by Coleman \cite{Coleman:1988cy} as part of a mechanism to restore quantum coherence in the context of wormhole physics. Using some of the algebraic structures of   $G$-TQFT2 developed in the context of open-closed topological string theory \cite{MooreSegal}, we find that regarding the centre of the group algebra of $G$, denoted  $ \cZ ( \mC( G ) ) \equiv \cH$, as a quantum mechanical Hilbert space gives a useful way to think about the one-dimensional topological quantum mechanics underlying $G$-TQFT2 and its two-dimensional geometrical structures.  The integer ratios ${ |G| \over d_R }
 $ play a central role in this discussion. Denoting as $P_R$ the projector basis elements of $ \cZ ( \mC( G ) )$, there is a handle-creation operator 
 \bea 
 \Pi = \sum_{R}  \left ({|G|\over d_R }\right )^2  P_R 
 \eea 
 which can also be expressed in terms of the structure constants of $ \cZ ( \mC( G ) )$ \cite{EFR,FHK}. 
 
 Considering the sums of $G$-TQFT2 amplitudes over all genera which define $G$-CTST as
  in \cite{MarMax,GardMeg} we   investigate the  stringy property of $S$-duality. 
We then   study  the finiteness properties of  $G$-TQFT2/$G$-CTST and their physical implications.  This  leads to the definition of  an inner product  for a polynomial algebra of surfaces, where the null states capture the finiteness relations. This  draws on the study of giant gravitons \cite{mst,gmt,HHI} in the context of AdS5/CFT4, notably features such as the departure from large $N$ factorization and their connection to finiteness and the holographic map for large operators \cite{BBNS,CJR}. As we explain,  $G$-TQFT2 amplitudes play a mathematical  role analogous to trace-observables of CFT4. This  leads to the consideration of a 2D/3D holographic duality involving  $G$-TQFT2. The discussion gives a  new perspective on the factorization puzzle associated with 3D wormholes in 2D/3D holography \cite{MaldaMaoz}.

The paper is organized as follows. In Section \ref{sec:Construct} we explain how the integer ratios $  ({ |G| \over d_R })^2  $ are constructed from amplitudes of $G$-TQFT2. In section \ref{sec:characters}  we generalize the discussion to show  how to construct the normalized characters of finite groups from boundary amplitudes of $G$-TQFT2. We explain the relation of this construction to existing algorithms for characters.  The normalized characters are defined as 
${ \chi^R ( g ) |\cC | \over d_R } $, where  $\chi^R ( g )$ is the character of a group element $g$ in the  
irrep $R$, $d_R$ is the dimension of the   irrep, and $ |\cC|$ is the number of elements in the conjugacy class $ \cC$ containing $g$. In Section \ref{sec:probabilities}, we describe probability distributions associated with the interpretation of  $G$-TQFT2 in terms of a 2D path integral, and the structure of the amplitudes as sums over irreps.  In Section \ref{Sdual} we describe an $S$-duality transformation on closed string amplitudes of $G$-CTST. While the expansion of $ G$-CTST at positive powers of the string coupling $g_{ st}$ is given in terms of positive powers sums of $ {|G|/d_R } $, the $S$-dual expansion is 
in terms of positive power sums of $ d_R$.  We give a geometrical interpretation of these positive power sums in terms of $G$-TQFT2 amplitudes for  entangled disconnected surfaces, where the entanglement is defined using 
projectors $P_R$ for the irreps living in the centre $ \cZ( \mC( G ))$ of the group algebra of $G$. In this section we also describe 
 the singularity structure of stringy partition functions  of $G$-CTST amplitudes as a function of the string coupling, exhibiting an interesting link between poles and residues of the partition functions and representation theoretic data.   Section \ref{sec:finiteK} is a detailed discussion of the implications of the finiteness of $G$ for relations between string amplitudes at different genera. This discussion leads to the introduction of a coupling between TQFT2 for $G$ and TQFT2 for symmetric groups, which we refer to as $\mC(G) \times ( \mC( S))_{ \infty } $-TQFT2. Powers of the handle-creation operator $\Pi$ play an important role in this coupling. Section \ref{sec:FNS3D} discusses the possibility of a  3D holographic interpretation for $\mC(G) \times ( \mC( S))_{ \infty } $-TQFT2 and in that scenario discusses the factorization puzzle associated with wormholes in 2D/3D holography \cite{MaldaMaoz}.

\section{  Constructing integer ratios $|G|/d_R$ from group products associated with surfaces   }\label{sec:Construct}

In general, group representation theory of a finite group $G$ over the complex numbers $ \mC $ is not a purely combinatoric subject. It can involve the solution of  eigenvalue equations with roots that may not be integer. Many interesting aspects of irreducible representations nevertheless have integrality properties. As we mentioned in the introduction, a striking property is that the dimension of every irrep is a divisor of the order of the group. The group multiplication table is a discrete and finite object. It is natural to ask if there is a  simple way to go from the group multiplication table to the integer ratios ${ |G|  \over d_R }$ for any group $G$ while working purely with integers. We  show in this section that this is indeed possible and that it involves group multiplications chosen to be of forms determined by two dimensional surfaces, and that the ratios are reconstructed by collecting the amplitudes of $G$-CTST over different genera and performing integer operations on this data.

Take a group $G$. Let $R$ be a label for its irreps, $d_R$ the dimension of the irrep.  We have the following well-known properties:
\bea 
\hbox{ Number of irreps}= \sum_R d_R^0 = \hbox{ Number of conjugacy classes, } 
\eea
while the sum of squares of the dimensions is 
\bea 
\sum_{ R } d_R^{ 2} = |G| 
\eea

In $G$-TQFT2, defined in terms of a sum of equivalence classes of $G$-bundles, weighted with inverse automorphism, the following equality is known \cite{DW,FQ,FHK} 
\bea\label{knownEq} 
&& \sum_{ R }  \left ( { |G| \over d_R   } \right )^{  2h  - 2  }  \cr 
&&  = \hbox{ Number of flat G-bundles on surface of genus $G$ counted with inverse automorphism } \cr 
&& = { 1 \over |G| } \sum_{ g_1 , g_2 , \cdots , g_{2h-1} , g_{ 2h} \in G } 
   \delta ( [g_1,g_2] [g_3,g_4]  \cdots  [ g_{2h-1} , g_{ 2h }]  ) 
\eea
where $ [g_1 , g_2] = g_1 g_2 g_1^{-1}g_2^{-1}$.
This equation has also been studied  in the mathematical  literature on finite groups and the geometry of surfaces\cite{Jones,Mednykh}.  

This follows using  Schur's orthogonality relations for matrix elements in irreducible representations. 
For any finite group $G$, we have
\bea
\sum_{g_1,g_2\in G}\sum_{b,c,d=1}^{d_R}
D^R_{ ab} (g_1) D^R_{ bc} (g_2) D^R_{ cd} (g_1^{-1}) D^R_{ de} (g_2^{-1})
=\left({|G|\over d_R}\right)^2\delta_{ae}
\eea
Consequently
\bea
\sum_{g_1,g_2,\cdots ,g_{2h}\in G}
\chi_R([g_1,g_2]\cdots [g_{2h-1},g_{2h}])=\left({|G|\over d_R}\right)^{2h} d_R
\eea
and hence
\bea\label{Gpart} 
\sum_{g_1,g_2,\cdots ,g_{2h}\in G}\delta([g_1,g_2]\cdots [g_{2h-1},g_{2h}])
&=&{1\over |G|}\sum_R \sum_{g_1,g_2,\cdots ,g_{2h}} d_R\chi_R([g_1,g_2]\cdots [g_{2h-1},g_{2h}])\cr
&=&|G|\sum_R \left({|G|\over d_R}\right)^{2h-2} 
\eea
$Z_h$,  the genus $h$ partition function, is then given by 
\bea\label{Zhirrepflat} 
&& Z_h =  { 1 \over |G| } \sum_{ g_1 , g_2 , \cdots , g_{2h-1} , g_{ 2h} \in G } 
\delta ( [g_1,g_2] [g_3,g_4]  \cdots  [ g_{2h-1} , g_{ 2h }]  )\cr 
&&  =  \sum_{ R }  \left ( { |G| \over d_R   } \right )^{  2h - 2  }  
\eea
This can also be written as 
\bea\label{AutP}  
Z_h = \sum_{P} { 1\over | \Aut (P)| }
\eea
where $P$ is a flat $G$-bundle which can be identified with an equivalence class of tuples 
$ (g_1 , g_2 , g_3 , g_4 , \cdots , g_{2h-1} , g_{2h}  )$ obeying the condition 
\bea 
g_1 g_2 g_1^{-1} g_2^{-1} \cdots g_{2h-1} g_{2h} g_{2h-1}^{-1} g_{2h}^{-1} = id. 
\eea
using the equivalence relation 
\bea 
( g_1 , g_2 , \cdots , g_{2h-1} , g_{2h} ) 
\sim ( g g_1 g^{-1} , g g_2 g^{ -1} , \cdots , g g_{2h-1} g^{-1} , g g_{2h} g^{-1} )  ~~ \hbox{for all } ~~ g \in G 
\eea
$ \Aut ( P) $ is the subgroup which leaves the tuple fixed and is an automorphism of the flat $G$-bundle $P$, $|\Aut (P)|$ is the order of the group.  For any group $G$, the ratio $ |G|/d_R $ is known to be  an integer.  A proof based on properties of algebraic integers is given in \cite{SimonsBook}.
This means that while each equivalence class of G-bundles contributes, in general,  a rational number  to 
the sum \eqref{AutP} the whole sum is an integer. This integrality is not obvious from a  topological point of view. This is discussed in \cite{Turaev07}.

\subsection{ Combinatoric construction  of $ { |G|^2  \over d_R^2 } $ }\label{CombConsGdR}  

Many interesting integral quantities in representation theory have combinatoric constructions. 
Examples include the dimensions of symmetric group irreps and the Littlewood-Richardson coefficients. 
For a general discussion of such problems see \cite{Stanley,BarceloRam}, 
For Kronecker coefficients of symmetric groups, a  lattice construction based on  ribbon graphs and 
 integer matrices arising from permutation  group multiplications  was recently given \cite{QMRibb}. 
Here we show that the  partition functions of $G$-TQFT2  on genus $h$ surfaces also allow us to 
construct the integers ${|G|^2\over d_R^2} \equiv a_R^2 $ by using 

\begin{itemize} 

\item Group multiplications of shape defined by the fundamental groups of surfaces. 

\item Searching among divisors of  integers. 

\end{itemize}

From \eqref{Zhirrepflat},  the genus one partition function, $Z_1$, is the number of conjugacy classes, which we will denote as $K$. 
The number of power sums we need to construct the set of all $ { |G| \over d_R } $  is $ K $. This means we need $Z_{ 2 } , Z_3 , \cdots , Z_{K+1 }$. 
Using Newton's identities, we get a polynomial of degree $ Z_1 = K$. 

It is convenient to define a matrix $X $ 
\bea 
X = {\rm Diag }  ( a_1^2 , a_2^2 , \cdots , a_K^2  ) 
\eea
Using \eqref{Zhirrepflat} we have 
\bea\label{tft2dat} 
&& \sum_{ R } a_R^2  = Z_2  = \tr X  \cr 
&& \sum_{ R } a_R^4  = Z_3 = \tr X^2  \cr 
&& \vdots \cr 
&& \sum_{ R } a_R^{ 2 K}= Z_{ K +1 }  = \tr X^{ K } 
\eea  
We use Newton's identities to convert these to elementary symmetric functions. 
In language familiar from the AdS/CFT treatment of branes, consider 
\bea\label{defFXx} 
&& F ( X , x ) = \det  ( x - X   ) = ( x - a_1^2    ) ( x  - a_2^2  ) \cdots ( x - a_K^2  ) \cr 
&& = x^K  - ( \tr X ) x^{ K -1} + { 1 \over   2 } ( ( \tr X )^2  - \tr X^2 ) x^{ n-2} + \cdots  + (-1)^K ( \det X )   \cr 
&& = x^K  - e_1 ( X ) x^{ K -1} + e_2 ( X ) x^{ K -2} + \cdots + (-1)^K e_K ( X ) 
\eea
The elementary symmetric functions are 
\bea 
&& e_0 ( X ) = 1 \cr 
&& e_1 ( X  ) = \sum_{ i } X_{ i }   \cr 
&& e_2( X ) = \sum_{ 1 \le i < j \le  K } X_{ i } X_{ j }  \cr 
&& e_{ l } ( X ) = \sum_{ 1 \le i_1 <  i_2 <  \cdots <  i_l  \le K } X_{ i_1} X_{ i_2} \cdots X_{ i_l } 
\eea
It is also useful to define $ E_l ( X ) = (-1)^l e_l ( X ) $ which leads to 
\bea 
F ( X , x ) 
&& = x^n + E_1 ( X ) x^{ n-1} + E_2 ( X ) x^{ n-2} + \cdots + E_{ K-1} ( X ) x + E_{ K } ( X ) \cr 
&& = \sum_{l =0  }^{ K }  x^{ K   - l } E_{ l } ( X ) 
\eea

From  \eqref{tft2dat} the coefficients of the powers of $x$ are given in terms of 
$G$-TQFT2 partition functions. From \eqref{defFXx} $a_R^2$ are the zeroes of $F ( X , x )$, which is viewed as a polynomial in $x$ with coefficients constructed from $G$-TQFT2 partition functions as above.  
So to construct the $ a_R$ from group theoretic combinatoric data, we
need to solve the polynomial equation 
\bea 
F  ( X , x ) = 0 
\eea
The elementary symmetric functions can be expressed in terms of traces of $X$ as 
\bea\label{ekformula}  
&& e_{ k  } ( X )  = \sum_{ p \vdash k  } { ( -1)^{ k    - \sum_{ i } p_i } \over \prod_{ i } i^{ p_i } p_i! }  \prod_i ( \tr X^i )^{ p_i }  \cr 
&& =  \sum_{ p \vdash k  }  { ( -1)^{ k  - \sum_{ i } p_i }  \over \prod_{ i } i^{ p_i} p_i! }
 \prod_i ( Z_{ i +1}  )^{ p_i }
\eea
Here $p$ is a partition of $k$, with $p_i$ parts of length $i$, so that  $ \sum_{ i } i p_i = k $. 
\bea 
\det X && =  \sum_{ p \vdash K  } { ( -1)^{ K   - \sum_{ i } p_i }\over \prod_{ i } i^{ p_i } p_i!  } \prod_i ( \tr X^i )^{ p_i }  \cr 
&& =  \sum_{ p \vdash K  }  { ( -1)^{ K   - \sum_{ i } p_i }  \over \prod_{ i } i^{ p_i } p_i! } 
\prod_i ( Z_{ i +1}  )^{ p_i }
\eea

If we did not know that  the $a_1 , a_2 , \cdots , a_K $ are integers, we would have to solve 
complicated polynomial factoring algorithms to find them. However, there are  simpler algorithms using this integrality (discussed e.g. at \cite{IntegerSolsPoly}).  

The numbers $ ( a_1^2 , a_2^2  ,\cdots , a_K^2 )$ are divisors  of $ \det X $ since 
$ F ( X , x=0 ) = (-1)^K  \det X $. Let 
\bea 
\Div_0 =  \hbox{ Set of divisors of } (-1)^K  F ( X , x=0 ) 
\eea
Each of the $a_R^2$  is a divisor of $ (-1)^K F ( X , x=0 )  $, i.e. an element of $ \Div_0$. 
Next note that $ (-1)^K F ( X , x=1 ) = \prod_{ R } ( a_R^2 -1 ) $. Let $r_i$ be the roots of $ F ( X , x=1 )$. The $ a_i^2 $ are among the  $ ( r_i+1) $. 
\bea 
\Div_1 = \hbox{ Set of divisors of } F ( X , x=1 ) \hbox{  shifted  up by} ~ 1
\eea
In general 
\bea 
\Div_l  = \hbox{ Set of divisors of } F ( X , x=l  ) \hbox{  shifted  up by}  ~ l 
\eea

Each element in the list  $\{ a_1^2 , a_2^2 , \cdots, a_{K}^2 \}  $ is  in the intersection 
\bea\label{intersection}  
\Div_0 \cap \Div_1 \cap \Div_2 \cdots \cap  \Div_{ K -1} 
\eea
and the list satisfies 
\bea\label{corrprod}  
 \prod_R ( a_R^2 - l )  = (-1)^{ K - l }   F ( X ,  l  ) \cr 
\eea
for all $l \in \{ 0 , 1, \cdots , K -1 \}$. 

{ \bf Claim } We show that if we take $K$ of these Divisor sets, and impose the above conditions \eqref{corrprod}, 
we will uniquely determine the  list of $a_R^2$.

\begin{proof} 
$$ F ( X , x ) = \sum_{ k =0 }^{ K }   x^{ K - k } E_k ( X ) $$  

Note that 
\bea 
&& F ( X , 1 ) = 1 + E_1 ( X ) + E_2 ( X ) +  \cdots + E_{ K -1} ( X ) + E_K ( X ) \cr 
&& F ( X , 2 ) = 2^K   + 2^{ K - 1 }    E_1 (X  ) +2^{ K - 2 }    E_2 (X )   +       \cdots                +    2 E_{ K -1} ( X )    + E_{ K } ( X ) \cr 
&& F ( X , 3 ) = 3^K   + 3^{ K - 1 }    E_1 (X  ) +3^{ K - 2 }    E_2 (X )   +       \cdots                +    3 E_{ K -1} ( X )    + E_{ K } ( X ) \cr 
&&  ~~~~~~~~~ \vdots \cr 
&& F ( X , K-1 ) =( K -1)^{K}  + ( K -1)^{ K-1} E_1( X ) + \cdots + ( K-1)  E_{ K -1} ( X ) + E_{ K } ( X ) \cr 
&& 
\eea
Rewrite this as 
\bea
&& F ( X , 1 ) - F ( X , 0 ) -1  =  E_1 ( X ) + E_2 ( X ) +  \cdots + E_{ K -1} ( X ) \cr 
&& F( X , 2 ) - F ( X , 0 ) -   2^K =     2^{ K - 1 }    E_1 (X  ) +2^{ K - 2 }    E_2 (X )   +       \cdots                +    2 E_{ K -1} ( X ) \cr 
&&  F ( X , 3 ) - F ( X , 0 )  -  3^K =     3^{ K - 1 }    E_1 (X  ) +3^{ K - 2 }    E_2 (X )   +       \cdots                +    3 E_{ K -1} ( X )  \cr 
&&  ~~~~~~~~~ \vdots \cr 
&& F ( X , K-1 ) - F ( X , 0 )  - ( K -1)^{K}  =   ( K -1)^{ K-1} E_1( X ) + \cdots + ( K-1)  E_{ K -1} ( X )\cr 
&& 
\eea
Note that 
\bea 
\begin{pmatrix} F ( X , 1 ) - F ( X , 0 ) -1 \cr 
 F( X , 2 ) - F ( X , 0 ) -   2^K \cr \vdots \cr 
  F ( X , K-1 ) - F ( X , 0 )  - ( K -1)^{K}  
  \end{pmatrix} 
  = \begin{pmatrix} 1 & 1 & \cdots & 1 \cr 2^{ K -1} & 2^{ K -2} & \cdots  & 2 \cr \vdots \cr ( K -1)^{K-1}  & ( K -1)^{ K -2} & \cdots & ( K -1) \end{pmatrix}   \begin{pmatrix} E_{  1} ( X ) \cr  E_{ 2} ( X ) \cr \vdots \cr 
   E_{ K -1} ( X ) \end{pmatrix}  \cr 
   && 
\eea
This is a linear system of equations giving the $F$'s in terms of the $E$'s. The transformation matrix is a 
non-singular Van der Monde matrix, which means that once we have chosen the $a$'s to reproduce the correct 
$F ( X , 0 ) , F ( X , 1 ) , \cdots , F ( X , K -1 ) $, the $ E_1 , \cdots , E_{ K -1} ( X )$ are also reproduced.

\end{proof} 

This completes the construction of the ratios $ { |G| \over d_R } $ from the combinatoric data of $G$-TQFT2 amplitudes. A further interesting question is whether equation \eqref{intersection} alone is sufficient (without using \eqref{corrprod}) to determine the  set of $a_{R}$. At the moment, as described above, one must determine the elements belonging to the intersection and then use \eqref{corrprod} to check that they are correct. Below, we perform some numerical experiments for $G = S_{n}$ in which we study only the intersection of the sets of divisors to determine the optimal $l$ necessary to identify the correct $a_{R}$. For a given $n$, we construct $F(X,0)$. Then we factorize $F(X,0)$ into its prime factors. For example, for $n=3$, $F(X,0) = 11664$, which factorizes into $2^{4} \times 3^{6}$. Then we use these prime factors, together with the corresponding exponents to build all possible divisors of $F(X,0)$. In the example above, $2^{e_{2}} \times 3^{e_{3}}$ with $ e_{2} = \{ 0,1,2,3,4 \}, e_{3} = \{ 0,1,\cdots ,6 \}$ will generate all possible divisors. To find all possible divisors of $F(X,1)$, we subtract 1 from the number we build and check if it is a divisor of $F(X,1)$. We follow a similar process to check if the number we build is a divisor of $F( X , 2 ) , F( X , 3 )$ etc. We keep only the elements that are divisors of the set $\{ F( X , 0 ) , F( X , 1 ) , \cdots , F( X , l ) \}$. In this way we construct the intersection \eqref{intersection} but for arbitrary $l$. We then check if this generates the correct set of $a_{R}$. The smallest $l$ value for which this occurs is denoted by $l_{opt}$. We performed these calculations for $S_{n}$ for $n = 3$ up to $n = 10$. 
The results are shown in Table \ref{Tab:Optl}. 

\begin{table}[h!]
  \begin{center}
  \begin{tabular}{|c|c|c|}
  \hline
  $n$ & $K$ & $l_{opt}$ \\
  \hline
  3 & 3 & 6 \\
  \hline	
  4 & 5 & 6 \\
  \hline
  5 & 7 & 7 \\
  \hline
  6 & 11 & 9 \\
  \hline
  7 & 15 & 7 \\
  \hline
  8 & 22 & 36 \\
  \hline
  9 & 30 & 36 \\
  \hline
  10 & 42 & 25 \\
  \hline
  \end{tabular}
      \caption{{\small {Table showing data for the optimal number of divisor sets to correctly determine the set of $a_{R}$ for $S_{n}$. The table shows results for $n = 3$ up to $n = 10$.}}}
  \label{Tab:Optl}
    \end{center}
\end{table}

We note that for $S_{n}$ it is indeed possible to construct the $| G | / d_{R}$ at least for small values of $n$ just by studying the intersection of the divisor sets of the $F(X,l)$. We further note that, for the $n$ values considered, the optimal number of divisor sets $l$ is always less than $2K$. 

Finally, we note that the power sums on the LHS of \eqref{knownEq}  have an interpretation in terms of string theory for a target space which is a disjoint union of points \cite{Sharpe2006,Sharpe2019,Sharpe2021}, which implies that the algorithm for converting the power sums to the integers $ ( { |G| \over d_R } )^2 $ we have described  has an interpretation as a construction of the string amplitudes for a target space point from a finite set of string amplitudes for the disjoint union \footnote{ We thank Eric Sharpe for bringing these papers to our attention and for discussions on this point}.

\section{ Constructing normalized characters  from group words and surfaces  } 
\label{sec:characters} 

Having shown how closed string amplitudes in $G$-TQFT2/$G$-CTST are used to construct the integer ratios $|G|/d_R$ and hence the $d_R$, we now consider the construction of characters $ \chi^R ( g )$. It turns out that the quantities appearing most directly from $G$-CTST amplitudes are an appropriately normalized form of the characters. Let $\cC_p$ be a conjugacy class of $G$ and let $|\cC_p| $ be the number of elements in the conjugacy class. We will denote by $T_p$ the sum of group elements, in the group algebra $ \mC ( G )$ 
\bea 
T_p = \sum_{ g \in \cC_p } g 
\eea
$T_p$ is a central element of $ \mC ( G )$, i.e. commutes with all elements in $\mC ( G )$. 
The normalized characters for $ \cC_p$ are, for $ g \in \cC_p$,  
\bea 
{ |\cC_p | \chi^R ( g ) \over d_R }  = { \chi^R ( T_p) \over d_R } 
\eea

The set of $T_p$ for all conjugacy classes spans the centre
  $\cZ(  \mC ( G )) $. Another basis for $ \cZ ( \mC ( G )) $ is given by the projectors labelled by irreducible representations $R$ 
  \bea 
  P_R = {d_R \over |G| } \sum_{ g } \chi^R ( g ) g^{ -1}  
  \eea 
The relation between the two bases is a Fourier transformation and plays an important role in this section. We will start in Section \ref{consnc} by giving the key formulae relating amplitudes of $G$-TQFT2 with boundaries to normalized characters, and then describe how these lead to a combinatoric construction of the characters. Since we restrict the discussion to combinatoric construction involving group multiplications and investigation of integer roots of polynomials, the output is the construction of the set of rational normalized characters (which are also integer as we will see) along with a polynomial with integer coefficients characterizing the non-rational characters which live in finite extensions of the rational numbers. In Section \ref{Burnside} we review a standard algorithm used 
for the construction of characters. In Section \ref{CharFT} we explain the link between the standard construction and the  discussion of Section \ref{consnc}.  The link originates from the fact that the higher genus amplitudes used in \ref{consnc} come from gluing 3-holed spheres and makes use of the Fourier transform on $ \cZ ( \mC ( G ) $.

\subsection{Construction of normalized characters from higher genus surfaces with boundary  }\label{consnc} 
A standard result in $G$-TQFT2 is that the amplitude for a  genus  $h$ surface with $r$ distinct boundaries, where the 
group element at the boundary is constrained to be in a conjugacy class $ \cC_p$ is 
\bea\label{bdyamps}  
&&  \sum_{ R } \left (  { |G| \over d_R } \right )^{ 2h - 2}     \left ( { \chi_R ( T_p) \over d_R } \right )^r  
\cr
 && =  \tr \left ( X^{ 2h -2 }  X_p^r  \right ) \cr 
&& = { 1 \over |G| } \sum_{  s_i , t_i } \sum_{ \s_1 . \cdots , \s_r \in \cC_p } 
 \delta \left ( ( \prod_{ i =1}^h s_i t_i s_i^{ -1} t_i^{ -1}  ) \sigma_1 \cdots \sigma_r \right ) 
\eea
$X$ is a diagonal matrix with diagonal entries equal to $ { |G| \over  d_R } $.  We have defined 
 $ X_p $ to be the diagonal matrix with matrix entries $  \left ( { \chi_R ( T_p) \over d_R } \right )$. 
Fixing $ h =1$,  we get power sums of the normalised  character from combinatoric data. 
\bea\label{powersumXp}  
 \tr ( X_p^r ) &=& \sum_{R} \left ( { \chi_R ( T_p) \over d_R } \right )^r  \cr
&=& { 1 \over |G| } \sum_{  s_1 , t_1 } \sum_{ \s_1 . \cdots , \s_r \in \cC_p } 
 \delta ( s_1 t_1 s_1^{ -1} t_1^{ -1}  \sigma_1 \cdots \sigma_r ) 
\eea

This gives the  combinatoric data reproducing $ \tr (X_p^r)$. It is the counting of $G$ bundles on genus one surfaces 
with $r$  punctures where the monodromy around each  puncture has the specified conjugacy class. 
In this way we can construct the characters of all conjugacy classes, using the same algorithm as in section \ref{sec:Construct}. 
The problem reduces to solving the polynomial equation
\bea 
F(X_p,x)=\det (X_p-x)=0 
\eea
As in Section \ref{CombConsGdR}, we can find integer solutions by considering the integer factors of 
$ F ( X_p , x=0) , F ( X_p , x=1) , \cdots $.  By factoring out the integer factors from $ F ( X_p , x )$ 
we are left with a polynomial $ P_{ \cC_p }  (  x )$ for every conjugacy class of a group $G$. 
For every group we can define a polynomial $ P_{ G } ( x )$ which is obtained by taking the product 
over all the conjugacy classes 
\bea 
\prod_{p} P_{ \cC_p } (x ) 
\eea 
and clearing all multiplicities, i.e. replacing any factor $p(x)^m$ for $ m >1 $ by $ p(x)$. 
This polynomial $P_{ G } ( x )$ is tabulated for non-Abelian  groups with order up to $60$ in Appendix \ref{sec:NAB60}. We have produced these polynomials from known character tables but they are in principle constructible using group multiplications shaped by surfaces with boundaries using \eqref{powersumXp}. 
 Note that for symmetric groups, these polynomials are always one. This is due to the fact that the normalized characters for symmetric groups are all integers, a fact which  was useful in the ribbon graph 
lattice algorithm for Kronecker coefficients given recently in \cite{QMRibb}.

The normalized characters $ { \chi^R ( T_p ) \over d_R } $ are algebraic integers, i.e. roots of a polynomial of the form 
$ x^n + a_{ 1} x^{ n-1} + \cdots + a_n $, where the leading coefficient $1$ and the other 
coefficients $a_i$  are integers. This is a consequence of the fact that 
they are eigenvalues of an integer matrix $ ( C_p)_q^{~~r}  = C_{ pq}^{~~r}  $  of structure constants of multiplication $T_p T_q = C_{pq}^{~~r}  T_r  $. This implies that the traces  $ \tr X_p^r $,  which are power sums of 
the normalized characters are integers.  Note that
the RHS of \eqref{bdyamps} and \eqref{powersumXp} can be expressed as a sum over equivalence classes of tuples of group elements 
satisfying the delta function condition. Each equivalence class is a flat $G$-bundle and has an automorphism group, consisting of group elements $g$ which fix the tuple, when acting by conjugation. In the sum, these equivalence classes are weighted by the inverse order of the automorphism group, so these are in general rational  numbers. Nevertheless, the sums in \eqref{bdyamps}
and \eqref{powersumXp} are integer due to the integrality of $ \tr X_p^r$. This seems to be an interesting, not a priori obvious, property  of $G$-bundles. 

It is useful to note that there are known  general finite  algorithms  \cite{BabRon} which produce all the matrix elements of irreducible representations over $ \mC$. They work by going over to the cyclotomic field $ Q_e$, where $e$ is the exponent of the group, i.e. the smallest positive integer such that $g^e$ is the identity for all the group elements. 
Our discussion above works for each fixed conjugacy class and produces the integer normalized characters along with a polynomial $ \cP_{ \cC_p } ( x )$ which defines an algebraic extension of the rational numbers containing the non-integer normalized characters for that conjugacy class. The non-integer roots will also live 
in the field  $ Q_{ e_p } $  where $e_p$ is the  smallest positive integer with the property that 
$ g^{ e_p} $ is the identity for $g$ in the conjugacy class $ \cC_p$. Typically $e_p $ will be much smaller than $e$.

For the symmetric group, the exponent is
\bea 
\prod_{ p \le n } p^{ \lfloor \ln ( n ) / \ln ( p ) \rfloor } 
\eea
For a short proof, see \cite{Symm-group-exp}. 
However it is known that the normalized characters are rational, for example by the Murnaghan-Nakayama construction \cite{FultonHarris}.  It is a useful 
fact  that if a normalized character is 
rational, it must be integer. 
The algorithm  for normalized characters we have described will determine all characters that can be expressed without
field extensions of the rationals. 
For a general group,  once we the integer characters are determined, we are left with the characters which require an 
extension of the rationals. It is worth emphasizing that our discussion in this paper 
is not focused on finding alternative efficient computations of characters to known methods, but to describe combinatoric  algorithms that go from group multiplications, of forms determined by two-dimensional surfaces, equivalently from amplitudes of combinatoric topological string theory, to the characters. By this route, we get to all the integer characters for any group and to polynomials $\cP_{\cC_p},\cP_{G}$, which have leading term $1$ and integer coefficients, and which characterise the non-integer characters. It would be  interesting  to consider efficient algorithms and general theoretical characterisation  of these polynomials for different choices of $ G , \cC_p$. 

The discussion above has focused on $h=1$.
If we use $ h =0$, we have  instead 
\bea 
 && \sum_{ R } { d_R^2 \over |G|^2 }  \left ( { \chi_R ( T_p) \over d_R } \right )^r  
 =  \tr \left ( X^{ -2 }  X_p^r  \right )  = { 1 \over |G| } \sum_{ \s_1 . \cdots , \s_r \in T_p } 
 \delta ( \sigma_1 \cdots \sigma_r ) 
\eea
An interesting problem  is to devise algorithms which take these
weighted power sums of $\left( {\chi_R (T_p) \over d_R}\right)$ and produce, as output, the normalized characters. It is intriguing that the $h=1$ data seems to lend itself to known algorithms we have used above, while the $h=0$ case seems less obvious. 

\subsection{ Link to Burnside's construction }\label{Burnside}  

Consider the sum
\bea
{1\over |G|}\sum_{s,t\in G}\sum_{\sigma\in T_p}\,\delta(sts^{-1}t^{-1}\sigma)\label{Genus1Sum}
\eea
The geometric interpretation of this sum is that it  gives the number of flat $G$-bundles on a torus with a single hole, counted
with inverse automorphism. The hole is in state $T_p$.
Recall that $T_p$ is the sum of all elements in conjugacy class ${\cal C}_p$ of $G$.We can replace the sum over $s$ by a sum over conjugacy classes $\cC_q$. 
The sum over $s$ above replaces $t$ with $T_q$ where $t\in {\cal C}_q$.
There is also a factor of $|G|/|T_q|\equiv |{\rm Sym} (\cC_q)| $ : $\Sym ( \cC_q )$ is the subgroup of $ G$ which commutes with $g \in \cC_q$, and $|\Sym ( \cC_q )|$ is the order of this subgroup.
The sum over $t$ then replaces $t^{-1}$ with $T_{q'}$ where $t^{-1}\in {\cal C}_{q'}$ and the sum over
$\sigma$ replaces $\sigma$ with $T_p$, so that
\bea
{1\over |G|}\sum_{s,t\in G}\sum_{\sigma\in T_p}\,\delta(sts^{-1}t^{-1}\sigma)
=\sum_{ q } {1\over |{T}_q|}\delta (T_q T_{q'}T_p)
\eea
The right hand side of the last line above has an intuitive geometrical interpretation: it is the partition function of a three
holed sphere with the hole in state $T_q$ glued to the hole in state $T_{q'}$.
The third hole is in state $T_p$.
The delta function is only non-zero when the product of the class functions $T_q$, $T_{q'}$ and $T_p$ multiply to give the
identity, with some multiplicity.
The identity always sits in a conjugacy class of its own.

As we have discussed in detail above, the sum (\ref{Genus1Sum}) produces a sum over the normalized characters.
The last equality above demonstrates that the sums defined by the TQFT are naturally related to the class algebra.
This connection has a natural counterpart in constructions of characters starting from the class 
algebra \cite{Burnside,Dixon,Schneider}.
Since these known mathematical algorithms are clearly closely related to the construction of characters from TQFT, 
it is worth reviewing them.

The normalized characters of the conjugacy classes
\bea
\omega_{Rp}={\chi_R(T_p)\over d_R}
\eea
obey an interesting algebra
\bea
\omega_{Rp}\omega_{Rq}=\sum_{r}n_{pqr}\omega_{Rr}\label{classalgebra}
\eea
The fusion coefficients $n_{pqr}$ are integers.
The algorithm of Burnside \cite{Burnside} constructs the characters using only algebra, assuming that the fusion coefficients
$n_{pqr}$ are known.
In practice the calculation of the $n_{pqr}$ can be carried out by multiplication in the group algebra.
The algorithm uses the class matrix $N_p$ defined by
\bea
(N_p)_{qr}=n_{pqr}
\eea
The class algebra (\ref{classalgebra}) implies that $\omega_{Rr}$ is an eigenvector of the class matrix.
Setting the identity class to be $r=1$, we see that 
\bea
 \omega_{R1}={\chi_R(1)\over d_R}=1
\eea
so that if we normalize the eigenvectors of the class matrix so that their first entry is 1, then the $p$th component of the
eigenvector is $\omega_{Rp}$.
To obtain the characters, we now need the dimensions $d_R=\chi_R(1)$ of each irrep.
By using character orthogonality it is easy to verify that the dimensions are fixed by the sum
\bea
\sum_r {\omega_{Rr}\omega_{Sr}\over |T_r|}={\delta_{RS}|G|\over \chi_R(1)\chi_S(1)}\label{fordim}
\eea
To summarize, Burnside's algorithm is
\begin{itemize}
\item[1.] Determine the conjugacy classes of $G$.
\item[2.] Compute $n_{rst}$ and hence the class matrices $N_r$.
\item[3.] Compute the eigenvalues of each $N_r$ normalized so that their first entry is 1.
This determines the $\omega_{Rp}$.
\item[4.] Determine the dimensions $d_R$ of the irreps using (\ref{fordim}).
\item[5.] Compute $\chi_R(T_p)=\omega_{Rp}d_R$.
\end{itemize}
Subsequent improvements of Burnside's algorithm were concerned with reducing the computational cost of step 3.
See \cite{Dixon,Schneider} for further details. 

\subsection{ Construction of characters and Fourier transform on the centre of $\mC ( G )$ }\label{CharFT}  

The centre of the group algebra $ \cZ ( \mC ( G ) )$ has two natural bases, related by a Fourier transform. 
The first basis $ \{ T_p \}$ corresponds to  conjugacy classes. The second basis set $ \{ P_R \}$ is labelled by irreps.  The projectors satisfy 
\bea 
P_R P_S = \delta_{ R S } P_S 
\eea
A useful property is 
\bea\label{eigTp} 
T_p P_R = { \chi^{ R } ( T_p) \over d_R } P_R 
\eea
The delta function on the group extends to $G$ and gives an inner product on $ \cZ( \mC ( G ) )$ 
\bea 
&& \delta ( T_p T_q ) = |T_q | \delta_{ p q' }  = |T_{ q'} | \delta_{ p q'} \cr \cr
&& = { |G|\over | \Sym (\cC_q ) |} \delta_{ p q' } =  { |G|\over | \Sym (\cC_{q'}  ) |} \delta_{ p q' } 
\eea
$\cC_{ q'}$ is the conjugacy class which contains the inverses  of  the group elements in the conjugacy class $ \cC_{ q} $. 
The inner product for the projectors is 
\bea 
\delta ( P_R P_S ) = \delta_{ RS} { d_R^2 \over |G| } 
\eea
The product in $ \cZ ( \mC ( G ))  $   in the $T_p$ basis is 
\bea 
T_p T_q = \sum_{ r }  C_{ pq}^{~~r} T_r = \sum_{ r }  { \delta ( T_p T_q T_{r'} ) \over | T_{ r } |} 
\eea
Consider the identity 
\bea 
\sum_{ q } { \delta ( T_q T_p^r T_{q'} ) \over |T_q |} = \sum_{ R } { |G| \delta ( P_R T_p^r P_R ) \over d_R^2 } 
\eea
which follows from taking the trace of $T_p^r$ in the two bases. We also have 
\bea 
\sum_{ q } { T_q T_{ q'} \over |T_q | } = { 1 \over |G| } \sum_{ g_1 , g_2 } g_1 g_2 g_1^{-1} g_2^{-1}  
\eea
and 
\bea 
\sum_{ R } { |G| \delta ( P_R T_p^r P_R ) \over d_R^2 }  = \sum_{ R } { \chi_R ( T_p^r  ) \over d_R } = \sum_{ R } 
\left ( { \chi_R ( T_p) \over d_R } \right )^r    
\eea
This leads to 
\bea 
{ 1 \over |G| } \sum_{ g_1  , g_2 \in G } \delta ( g_1 g_2 g_1^{ -1} g_2^{-1} T_p^r ) = \sum_{ R } \left ( { \chi_R ( T_p) \over d_R } \right )^r  =  \Tr (X_p^r)
\eea
We have thus recovered the identity  \eqref{powersumXp}  by taking the trace of $T_p^r$ in the two bases 
 for $ \cZ ( \mC ( G ) $.  The diagram in  \eqref{fig:Tp2tr} illustrates the geometrical nature of the calculation for  $r=2$ : 
\bea\label{fig:Tp2tr} 
\begin{gathered}
\includegraphics[width=0.075\columnwidth]{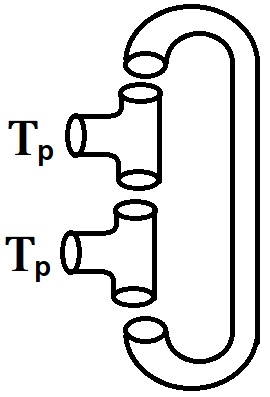}%
\end{gathered}
\longrightarrow
\sum_q\delta (T_{q'}T_p^2 T_q){1\over |T_q|}
=\sum_R\delta (P_RT_p^2 P_R){1\over\left({d_R^2\over |G|}\right)}
\eea
Note that 
\bea
\sum_q\delta (T_{q'}T_p^2 T_q){1\over |T_q|}
&=&\sum_{q,r'}\delta(T_{q'}T_p C^{~~r_1}_{pq}T_{r_1}){1\over |T_q|}\cr
&=&\sum_{q,r_1,r_2}C^{~~r_1}_{pq}C_{pr_1}^{~~r_2}\delta(T_{q_1}T_{r_2}){1\over |T_q|}\cr
&=&\sum_{q,r_1}(C_p)_q{}^{r_1}(C_p)_{r_1}{}^r\cr
&=&\Tr (\tilde{X}_p^2)
\eea
where $(\tilde{X}_p)_q^r=C_{pq}{}^r$. From \eqref{eigTp} we see  that the eigenvalues of $ \tilde X_p$ are nothing but the entries of the diagonal matrix $X_p$ we defined in Section \eqref{consnc}.  
 
The connection to Burnside's algorithm is that the coefficients $ C_{ pq}^{ ~~ r  }$ are equal to the $n_{ pqr}$ appearing in the product of the normalized characters \eqref{classalgebra}. The calculation of the eigenvalues of 
$ ( N_p )_{qr}  = n_{ p q r }  = (\tilde X_p )_q^r$ involves the calculation of 
\bea 
\det ( \tilde X_p - \lambda ) 
\eea
The coefficients of the powers of $ \lambda $ are elementary symmetric polynomials expressible in terms of 
power sums   $ \tr  \tilde X_p^r = \tr X_p^r $. 
These powers appear in
\bea
\det (\tilde{X}_p-\lambda)=\det (X_p-\lambda)
\eea
Our discussion in Section \ref{consnc} introduced the polynomial $\det (X_p-\lambda)$ as a tool to extract the 
diagonal entries of the matrix $ X_p$ from the power sums of these entries (a tool which we also used in Section
\ref{sec:Construct} to construct $ |G|/d_R$), whereas $\det (\tilde{X}_p-\lambda)$ arises in Burnside's construction 
from the diagonalization of the matrix of structre constants $ ( N_p )_{qr}  = C_{pq}{}^r $.

\section{Probability distributions from $G$-TQFT2 and $G$-CTST}\label{sec:probabilities}  

The formulation of $G$-TQFT2 as a  topological lattice gauge theory with plaquette weight enforcing the flatness condition can be viewed as a discrete  path integral in a two-dimensional theory, and realizes the axiomatic formulation of TQFT by Atiyah \cite{Atiyah}.  This approach to $G$-TQFT2 is a finite group version of \cite{Witten2dYM} which is known to be equivalent to the formulation in \cite{FHK}. 
 The state space associated with a circle boundary is the centre $\cZ( \mC(G))$ of the group algebra $ \mC( G ) $  of $ G$.  Using the path integral interpretation, it is natural to use the partition function for a $2$-manifold with for $n$ incoming circle boundaries  and $m$ outgoing circle boundaries to define probabilities for transitions between $ \cH^{ \otimes n } $ to $ \cH^{ \otimes m }$. As we will see the amplitudes are invariant under exchange of the states in the tensor factors, so we have transitions from the symmetrised product $ S^n  ( \cH )$ to $ S^m ( \cH)$. We will write explicit formulae for the  probabilities of such transitions within $G$-TQFT2 and  by summing over different genuses with weight $g_{st}^{ 2h-2}$ we will get analogous transition probabilities for $G$-CTST. The transition probabilities are squares of amplitudes. These amplitudes in turn have an interesting structure, containing a sum over a label $R$ for  irreducible representations which contribute according to some positive weights. These positive weights themselves define  probability distributions over the irrep labels. These probability distributions include the Plancherel distribution for finite groups \cite{BorOk} and generalizations thereof. Following  \cite{MarMax} the sum over $R$ formulae for the amplitudes  have an interpretation in terms of a classical ensemble and the irrep labels $R$  provide  an example of Coleman's  $\alpha$-eigenstates which were used to explain quantum coherence in the context of  wormholes \cite{Coleman:1988cy,Giddings:1988cx}. Building on known structures of $G$-TQFT2 \cite{MooreSegal}, we will review  how the sum over irrep sectors arises from the gluing relations of $G$-TQFT2.  These are based on the properties of the algebra $ \cZ( \mC( G  ) )= \cH$. The existence of two bases for $ \cH$, one labelled by conjugacy classes and one labelled by irreps plays a key role in understanding the sectors. To accommodate the in-out states of the general transition amplitudes between circles, and take advantage to the gluing relations, it is useful to consider quantum mechanics based on the state space 
\bea 
\bigoplus_{ n =0}^{ \infty } S^n ( \cH) 
\eea
built from the algebra $ \cH$. This may be viewed as a one-dimensional quantum system underlying the sum over $ \alpha$-states arising in the amplitudes of $G$-TQFT2.

\subsection{ Plancherel distribution for $G$ and geometrical generalizations in $G$-TQFT2 } \label{Planch}

The Plancherel distribution makes a natural appearance in the sphere partition function, which is given by the sum
\bea 
Z_{ S^2} = \sum_{ R } \left ( { d_R^2 \over |G|^2 } \right )  = { 1 \over |G| } 
\eea
The algorithms in Section 3 allow us to use the genus one and higher partition functions to
construct the ratios $ { |G| \over d_R } $. Using these ratios and the formula above, we can add up to recover $Z_{ S^2}$. In this sense the higher genus partition functions allow us to reconstruct the genus zero partition function. 

By multiplying with $ |G| $ we have 
\bea 
|G| Z_{ S^2} = \sum_{ R } \left ( { d_R^2 \over |G| } \right ) =1 
\eea
The summands are positive numbers which define the  Plancherel
distribution. We can give an interpretation of each probability in terms of disc partition functions. 
The disc partition function with boundary condition $g\in G$ is proportional to  the delta function on the group 
\bea\label{discPF}  
|G| Z_{ D^2} ( g ) = \delta ( g ) = \sum_{ R } { d_R \chi^R ( g ) \over |G| }
\eea
The disc partition function is defined for group algebra elements $ \sum_{ g } c_g g \in \mC ( G ) $ as
\bea 
Z_{ D^2} ( \sum_{ g } c_g  )  = \sum_{ g } c_g Z_{ D^2} ( g ) 
\eea
Consider the projector element in $ \mC ( G )$ associated with an irrep $R$
\bea 
P_R ={  d_R \over |G| } \sum_{ g } \chi_R ( g ) g^{-1} 
\eea
where $ \chi_R ( g )$ is the character of $g\in G $ in irrep $R$. 
\bea
Z_{ D^2} ( P_R )  = \sum_{ g } { d_R \chi^R ( g ) \over |G|  }  Z_{ D^2} ( g^{-1}  ) 
\eea
Using \eqref{discPF} we find
\bea 
|G| Z_{D^2}(P_R) = {d_R^2\over |G|} \label{NothingToDisc}
\eea
This reproduces the probabilities of the Plancherel distribution for a finite group, directly from a disc partition function. 

The $G$-TQFT2 perspective  on the Plancherel distribution shows that it has a number of  generalizations with a geometric interpretation. These generalizations are motivated by considering partition functions for surfaces with $h$ handles and $b$ boundaries.
The partition function is 
\bea 
Z_{h,b;T_{p_1},\cdots,T_{p_b}} & = & \sum_{R}\left({d_R\over |G|}\right)^{2-2h-b} \chi^R (T_{p_1})\cdots\chi^R(T_{p_b}) \cr 
& = &  { 1\over |G| } \delta ( \Pi^{ h -1 }  T_{ p_1} \cdots T_{ p_b})
\eea
where $\Pi$ is the handle-creation operator 
\bea
\Pi & = &  \sum_{R}  \left ( { |G | \over d_R } \right )^2 P_R \cr 
 & = & \sum_{ g_1 , g_2 \in G } 
 	g_1 g_2 g_1^{-1} g_2^{-1} = \sum_{p } T_p T_{p'} ~ |\Sym ( \cC_p ) |
\eea
This partition function is normalized so that it is consistent with natural geometrical gluing relations as we now explain.
Gluing is performed using the inverse cylinder and the boundary in state $T_p = \sum_{g \in \cC_p } g $ is glued to a boundary in state $T_p' = \sum_{ g \in \cC_p} g^{-1}$, and summing the label $p$
over all the conjugacy classes of $G$.
We can absorb the factors associated to the inverse of the cylinder in the definition
\bea
  T_p^*={T_{p'}\over |T_p|\,|G|}
\eea
 A basic identity that  can be  used to perform any gluing is
\bea
\sum_{p}\chi_R(AT_p)\chi_S(T_p^*B)=\sum_{p}{\chi_R(AT_p)\chi_S(T_{p'}B)\over |T_p|\, |G|}=\delta_{RS}\chi_R(AB)
\eea
where  $A$ and $B $ to belong to $ \mC( G)$ and at least one is in the centre $\cZ( \mC(G))$.  It is also useful to have the gluing equation in terms of projectors : 
\bea\label{glueR}  
\delta ( A B ) =\sum_{R}  \delta ( A P_R) \delta ( P_R B ) { |G| \over d_R }
\eea
where we also require $A$ and $B $ to belong to $ \mC( G)$ and at least one is in the centre $\cZ( \mC(G))$.  As an example of some gluing relations, gluing a disc to a $b$ holed sphere gives a sphere with $b-1$ holes
\bea
\sum_{p} Z_{D^2}(T_p)Z_{h,b;T_{p_1},\cdots,T_{p_{b-1}}T_p^*}
=Z_{h,b;T_{p_1},\cdots,T_{p_{b-1}}}
\eea
and gluing two holes increases the number of handles by 1 and decreases the number of boundaries by 2
\bea
\sum_{p} Z_{h,b;T_{p_1},\cdots,T_{p_{b-2}}T_pT_p^*}
=Z_{h+1,b-2;T_{p_1},\cdots,T_{p_{b-2}}}
\eea

As mentioned above, the Plancherel distribution admits a number of interesting generalizations.

We can consider  a $G$-TFT2 map from $n$ holes to $m$ holes.
Associate  $T_{p_1}\cdots T_{p_n}$ to the initial holes and $T_{q_1'}\cdots T_{q_m'}$ to the final holes.
The relevant partition function, for a surface with $h$ handles, is given by
\bea
Z_{h,m+n,T_{p_1}\cdots T_{p_n}T_{q_1}^*\cdots T_{q_m}^*}&=&
\sum_{R}\left({d_R\over |G|}\right)^{2-2h-m-n} \chi^R (T_{p_1})\cdots\chi^R(T_{p_n})
\chi^R (T_{q_1}^*)\cdots\chi^R(T_{q_m}^*)\cr\cr\cr
&=&
\begin{gathered}
\includegraphics[scale=0.4]{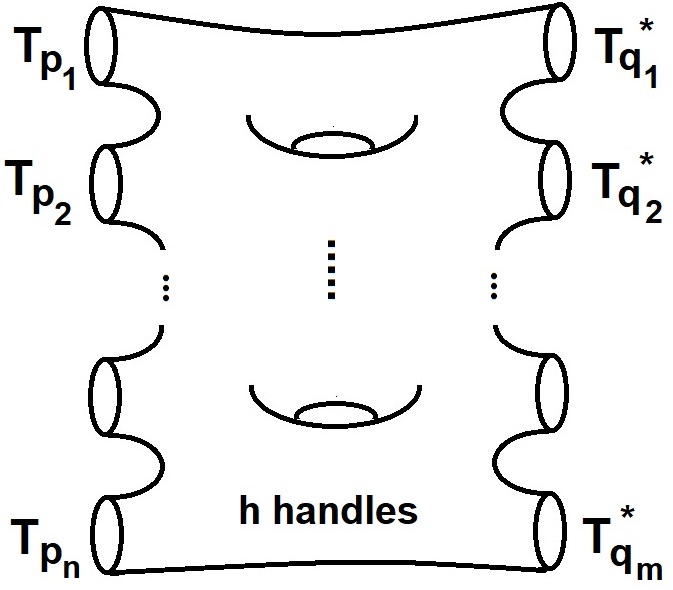}
\end{gathered}
\eea
This amplitude naturally defines a probability for a state of $n$ circles with boundary conditions $ T_{ p_1} , T_{ p_2}, \cdots , T_{ p_n}$ to evolve to a state of $m$ circles with boundary conditions $T_{q_1}^* , T_{ q_2}^* , \cdots T_{ q_m}^*$. 
 The associated probability is 
\bea
p(h,m+n,T_{p_1}\cdots T_{p_n}T_{q_1}^*\cdots T_{q_m}^*)
={Z_{h,m+n,T_{p_1}\cdots T_{p_m}T_{q_1}^*\cdots T_{q_m}^*}Z_{h,m+n,T_{q_1}\cdots T_{q_m}T_{p_1}^*\cdots T_{p_n}^*}\over N_{h,m,T_{p_1}\cdots T_{p_n}}}
\eea
These are indeed  positive and correctly normalized as we explain, allowing a probability interpretation. This follows  because $ \chi_R ( g^{-1} ) $ is 
the complex conjugate of $ \chi^R (g)$ for finite groups (where any representation can be made unitary): consequently 
$ \chi_R ( T_p^*) = { \overline { \chi_R ( T_p) }  \over |G| |T_p| } $
where the normalization is determined by
\bea
N_{h,m,T_{p_1}\cdots T_{p_n}}^2&=&
\begin{gathered}
\includegraphics[scale=0.4]{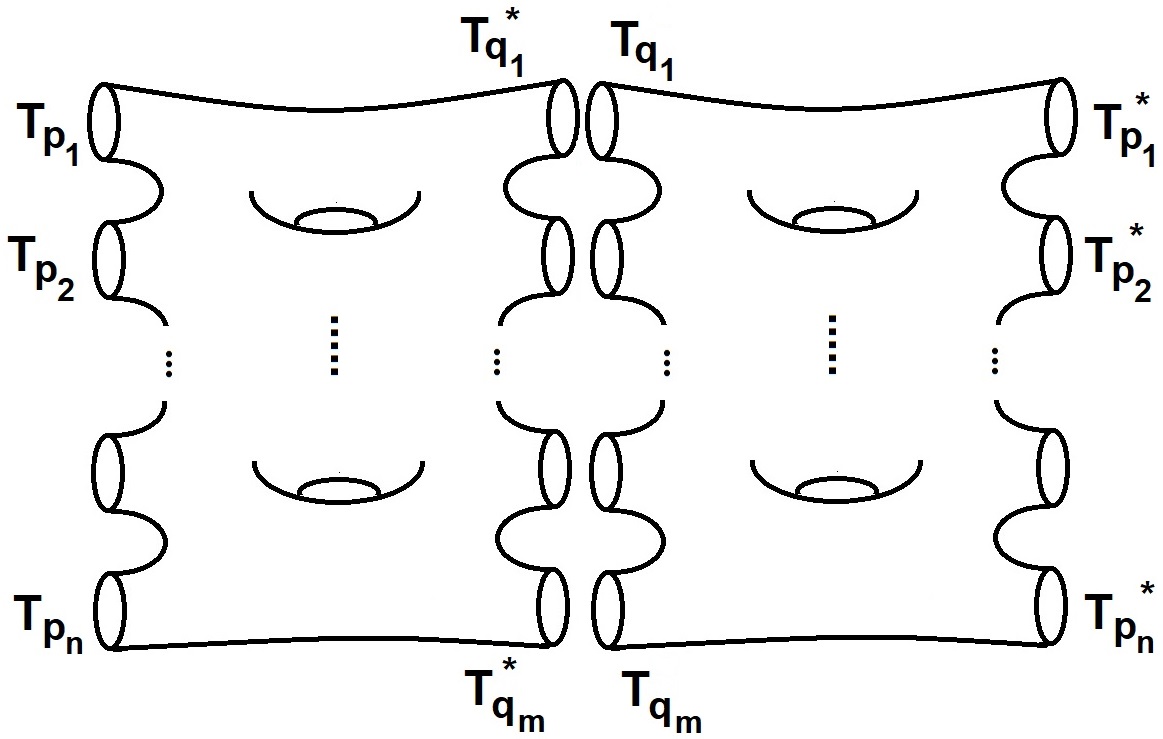}
\end{gathered}\cr\cr
&=&
\begin{gathered}
\includegraphics[scale=0.4]{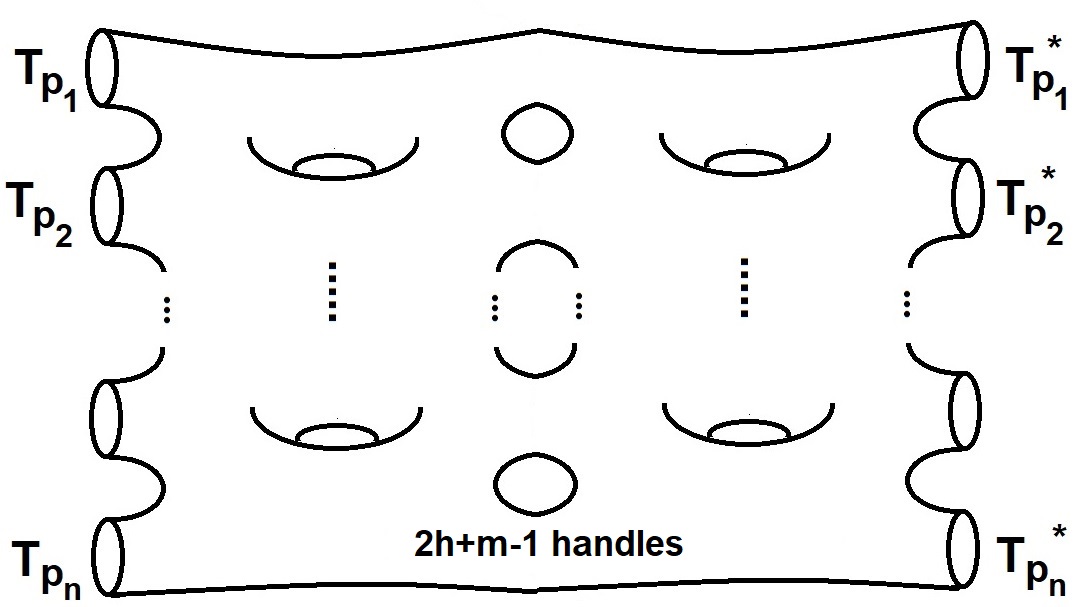}
\end{gathered}\cr\cr
&=&\sum_{n}\sum_{q_1,q_2,\cdots q_m}
Z_{h,m+n,T_{p_1}\cdots T_{p_n}T_{q_1}^*\cdots T_{q_m}^*}
Z_{h,m+n,T_{q_1}\cdots T_{q_m}T_{p_1*}\cdots T_{p_n*}}\cr
&&
\eea
We find that 
\bea 
N_{h,m,T_{p_1}\cdots T_{p_n}}^2 = Z_{ 2h + m -1 , 2n ;  T_{ p_1 } , \cdots , T_{ p_n } ;  T_{ p_1}^* , \cdots , T_{ p_n}^* } 
\eea
This follows by using  
\bea 
\sum_{ q } \chi_R ( T_q) \chi_S ( T_q^* ) = \delta_{RS} 
\eea
The construction of normalizing probabilities using higher genus partition functions has been employed in a conformal field theory context in \cite{copto} in connection with giant gravitons. $G$-TQFT2 provide a simpler realisation of the same concept. 

\subsection{$G$-TQFT2 partition functions, baby-universe operators and quantum mechanical state space   }\label{PlanchTwo}

We have considered how partition functions for surfaces with boundary define amplitudes and  probabilities for transitions between conjugacy class observables. These  amplitudes come from a sum over irreps $R$. The weights for these different $R$ can themselves be interpreted as defining a probability distribution. Following the discussion  in \cite{MarMax}, the  different $R$ can be identified with   the $\alpha$-states of Coleman \cite{Coleman:1988cy}.

\begin{figure}[h]%
\begin{center}
\includegraphics[width=0.35\columnwidth]{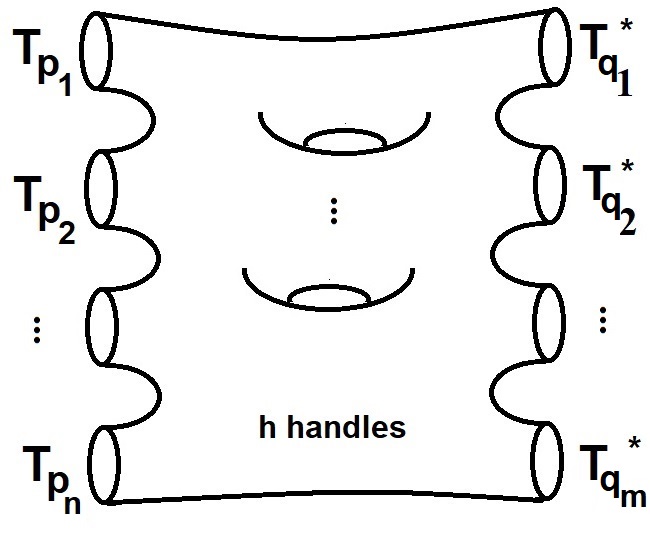}%
\caption{Genus $h-1$ surface with insertions of elements of the class algebra. A surface with $h=2$ is shown.}
\label{fig:PF}
\end{center}
\end{figure}

Consider the partition function, on a surface of genus $h$, with $n$ insertions of the form
\bea
{\chi_R(T_{p_i})\over d_R}\qquad\qquad i=1,2,\cdots,n
\eea
and $m$ insertions of the form
\bea
{\chi_R(T_{q_i}^*)\over d_R}\qquad\qquad i=1,2,\cdots,m
\eea
The partition function is given by
\bea 
Z_{\Sigma_{h,T_{p_1}\cdots T_{p_n}T_{q_1}^*\cdots T_{q_m}^*}}
=\sum_R \left({|G|^2\over d_R^2}\right)^{h-1}{\chi_R(T_{p_1})\over d_R}\cdots {\chi_R(T_{p_n})\over d_R}
{\chi_R(T_{q_1}^*)\over d_R}\cdots {\chi_R(T_{q_m^*}) \over d_R}\label{pfensemb}
\eea
This partition function can be represented as the path integral on the surface  shown in Figure \ref{fig:PF}.

By cutting the path integral open, we can get two states.
For example, the ket vector
\bea
|T_{p_1},T_{p_2},\cdots,T_{p_n}\rangle
=
\begin{gathered}
\includegraphics[width=0.18\columnwidth]{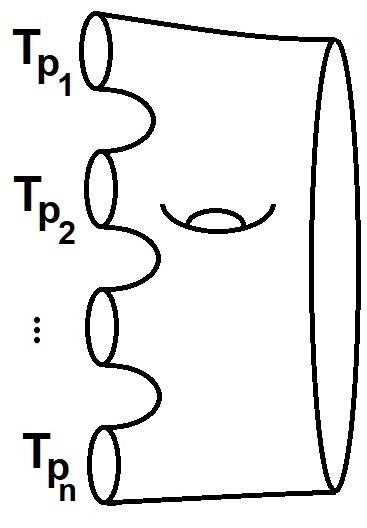}
\end{gathered}
\eea
belongs to $\Sym^k ({\cal H})$, the symmetric product of $k$ copies of ${\cal H}$.
The ket vector is invariant under swapping the $T_i$s so that for example
\bea
|T_{p_1},T_{p_2},T_{p_3},\cdots,T_{p_n }\rangle = |T_{p_2},T_{p_1},T_{p_3},\cdots,T_{p_n}\rangle
\label{ketsym}
\eea
This follows directly from the starting point (\ref{pfensemb}). 
We can also define the bra vector
\bea
\langle T_{q_1}^*,T_{q_2}^*,\cdots,T_{q_m}^*|=
\begin{gathered}
\includegraphics[width=0.18\columnwidth]{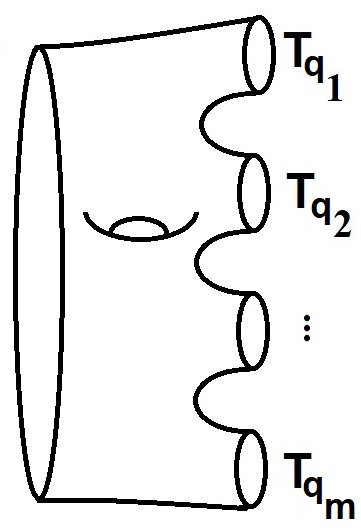}
\end{gathered}
\eea
which enjoys the same symmetry
\bea
\langle T_{q_1}^*,T_{q_2}^*,T_{q_3}^*,\cdots,T_{q_n}^*|=\langle T_{q_2}^*,T_{q_1}^*,T_{q_3}^*,\cdots,T_{q_m}^*|
\eea
The inner product of this bra and ket corresponds to the partition function
\bea
\langle T_{q_1}^*,T_{q_2}^*,\cdots,T_{q_m}^*|T_{p_1},T_{p_2},\cdots,T_{p_n}\rangle
=Z_{\Sigma_{h,T_{p_1}\cdots T_{p_n}T_{q_1}^*\cdots T_{q_m }^*}}
\eea
We can define a set of commuting operators, which act as follows
\bea\label{BabyUC} 
\hat{T}_a|T_{p_1},T_{p_2},\cdots,T_{p_n}\rangle=|T_{p_1},T_{p_2},\cdots,T_{p_n},T_a\rangle
\eea
These operators commute thanks to the symmetry (\ref{ketsym}). These are the baby universe creation operators in this model \cite{MarMax,Coleman:1988cy}. 
Any state in the Hilbert space can be obtained by acting on the ``vacuum state'' $|0\rangle$, which corresponds to the
surface without any boundary circles.
So, for example
\bea
|T_{p_1},T_{p_2},\cdots,T_{p_k}\rangle=\hat{T}_{p_1}\,\hat{T}_{p_2}\,\cdots\,\hat{T}_{p_k}|0\rangle
\eea
We also have
\bea
\langle T_{l_1}^*,T_{l_2}^*,\cdots,T_{l_k}^*|=\langle 0|\hat{T}_{l_1}^\dagger \,\hat{T}^\dagger_{l_2}\,\cdots\,
\hat{T}^\dagger_{l_k}
\eea
The $T_i$s and the $T_i^\dagger$s commute, which is again a direct consequence of (\ref{pfensemb}). 
This means that we can simultaneously diagonalize all of these operators.
The simultaneous eigenkets are $R$ states $|R\rangle$ which obey
\bea
\hat{T}_p |R\rangle ={\chi_R(T_{p})\over d_R}|R\rangle
\eea
i.e. the eigenvalues of the $\hat{T}_p$ operators are the normalized characters.
These states are the analog of the $\alpha$-eigenstates introduced in \cite{Coleman:1988cy}.
Similarly
\bea
\langle R|\hat{T}^\dagger_p  ={\chi_R(T_{p}^*)\over d_R} \langle R|
\eea

We will now show that the $G$-TQFT2 partition function can be interpreted as a sum over a classical ensemble of theories.
The norm of the vacuum state is given by
\bea
\langle 0|0\rangle =Z_{\Sigma_h}=\sum_R \left({|G|^2\over d_R^2}\right)^{h-1}\equiv \mathfrak{z}
\eea
To get a correctly normalized distribution we should divide by the norm of the vacuum state.
The partition function is

\bea\label{sumRsecs} 
Z_{\Sigma_{h,T_{p_1}\cdots T_{p_k}T^*_{q_1}\cdots T^*_{q_l}}}
&=&\langle \hat{T}^\dagger_{q_1}\cdots \hat{T}^\dagger_{q_l}\hat{T}_{p_1}\cdots \hat{T}_{p_k}\rangle\cr
&=&\sum_R \left({|G|^2\over d_R^2}\right)^{h-1}{\chi_R(T_{p_1})\over d_R}\cdots {\chi_R(T_{p_k})\over d_R}
{\chi_R(T_{q_1}^*)\over d_R}\cdots {\chi_R(T_{q_l}^*)\over d_R}\cr
&\equiv& \mathfrak{z}\sum_R p_R {\chi_R(T_{p_1})\over d_R}\cdots {\chi_R(T_{p_k})\over d_R}
{\chi_R(T_{q_1}^*)\over d_R}\cdots {\chi_R(T_{q_l}^*)\over d_R}
\eea
where in the last line we pulled out the normalization of the vacuum state and we have introduced the notation
\bea
p_R ( h ) ={\left({|G|^2\over d_R^2}\right)^{h-1}\over\sum_T \left({|G|^2\over d_T^2}\right)^{h-1}}\label{ThryProb}
\eea
It is obvious that
\bea
\sum_R p_R ( h )= 1
\eea
If we set $h=0$ we find that
\bea
p_R (h=0)={d_R^2\over |G|}
\eea
which is the  Plancherel measure.

Notice that we can write the partition function \eqref{sumRsecs} as follows
\bea 
\sum_{R} p_R \cO_{p_1 , p_2 , \cdots , p_l ; q_1 , \cdots , q_l  } ( R )
\eea
which illustrates the fact that we can interpret the partition function as computing observables in a classical ensemble of theories. This interpretation has been discussed in \cite{MarMax}.

From the  point of view discussed above  the weight  $p_R ( h=0)$ is the normalized version of the weight $ ({ |G|^2 \over d_R^2 })^{ h-1}  $ for the normalized characters in \eqref{sumRsecs}. We observed earlier that the Plancherel distribution can be understood in terms of the amplitude for transition from the vacumm to disc (\ref{NothingToDisc}). Generalizing this observation, the partition function for genus $h$ with one boundary and with insertion of the central element $P_R$ at the boundary is given by  
\bea 
Z_{ h , 1 , P_R  } = { 1\over |G|}
\delta ( \Pi^h P_R ) = \left (  { |G|^2 \over d_R^2} \right )^{ h-1}  
\eea 
We can use the gluing relation \eqref{glueR} to write the partition function as a sum over $R$-sectors 
\bea\label{RexpGeom} 
&& Z_{\Sigma_{h,T_{p_1}\cdots T_{p_k}T^*_{q_1}\cdots T^*_{q_l}}}
= { 1 \over |G | }
\delta ( \Pi^h T_{ p_1 } \cdots T_{ p_k }  T_{ q_1}^* \cdots T_{ q_l}^* ) \cr 
&& = \sum_{R} \delta ( \Pi^h P_R )
{ |G | \over d_R^2 }  \delta ( P_R  T_{ p_1 } \cdots T_{ p_k }  T_{ q_1}^* \cdots T_{ q_l}^*) \cr 
&& = \sum_{R} \left ( { 1 \over |G| } \delta ( \Pi^h P_R ) \right) 
{ |G |^2 \over d_R^2 }  \delta ( P_R  T_{ p_1 } \cdots T_{ p_k }  T_{ q_1}^* \cdots T_{ q_l}^*) 
\eea
This shows that the sum over $R$-sectors in \eqref{sumRsecs} can be interpreted geometrically by cutting the genus $h$ transition  surface from the in and out circles, along a circle which separates the handles from the in-out states. On this circle we insert a complete set of projectors $P_R$ which span $\cH= \cZ ( \mC(  G ))$. The equality \eqref{RexpGeom} is illustrated below :
\bea
\begin{gathered}
	\includegraphics[width=0.18\columnwidth]{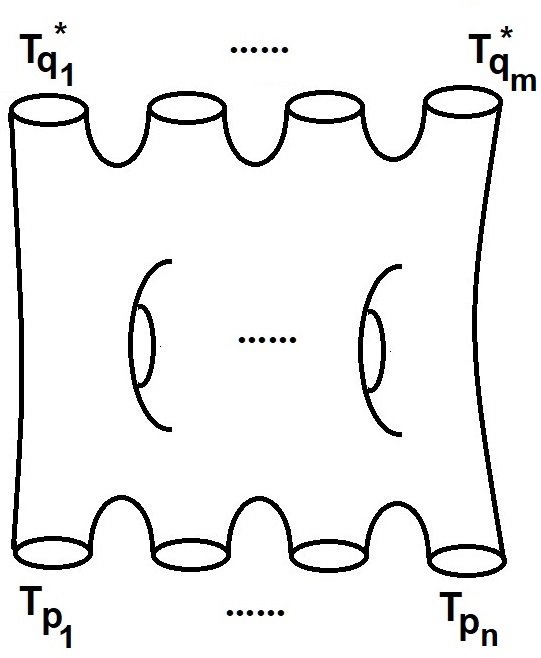}
\end{gathered}
=\sum_R \,\,\,
\begin{gathered}
	\includegraphics[width=0.26\columnwidth]{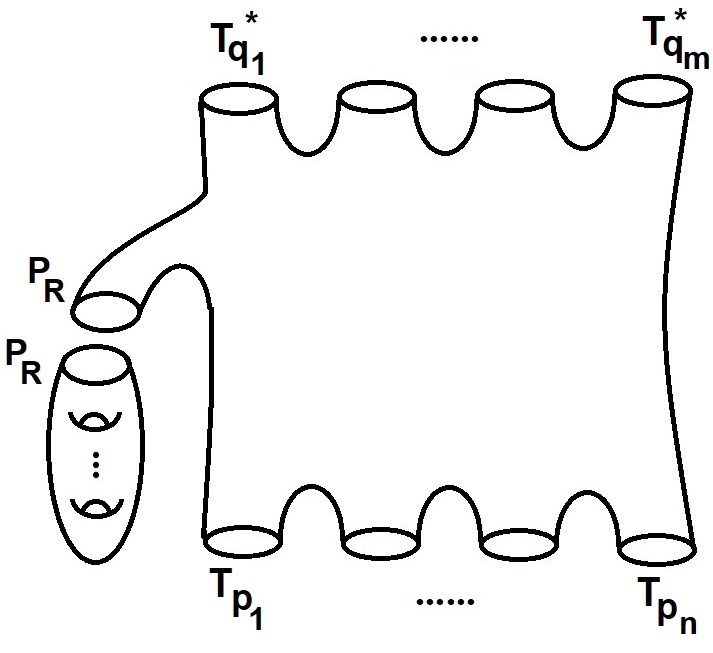}
\end{gathered}
\eea

It is useful to note that a central role is played in the above discussion by the algebra $ \cZ( \mC ( G))$. We can define a topological quantum mechanics by taking as state space $\cH =\cZ( \mC ( G)) $. This state space is a complex vector space, but equipped with additional structures. It has an associative product. There is a map $ \delta : \cH \rightarrow \mC$ which is defined by using the delta function on the group and extending to the group algebra (and its centre) by linearity. We can define an inner product 
\bea 
g  (  \sum_{g_1} a_{ g_1} g_1 , \sum_{g_2} b_{ g_2 } g_2 )
= \sum_{g_1 , g_2 } 
(a_{ g_1})^* b_{ g_2} \delta ( g_1  g_2^{-1} )
\eea
On the projector basis the inner product is 
\bea 
g ( P_R , P_S ) = \delta_{RS} {d_R^2 \over |G| } 
\eea
Thus the inner product is non-degenerate and $\cH$ is indeed a Hilbert space. The transition probabilities considered earlier in this section are maps from 
\bea 
S^n ( \cH) \rightarrow S^m ( \cH)
\eea 
defined using  the algebra  multiplication in $ \cH$ and the delta function map $\delta : \cH\rightarrow \cH$. The state spaces $ \cH$ and 
\bea 
\bigoplus_{ n=0}^{ \infty }  S^n ( \cH)  \equiv S^{ \infty } ( \cH) 
\eea
 provide the complete framework for computing the amplitudes and  probabilities considered and may  be usefully thought of as defining a one-dimensional topological quantum mechanical system encoding amplitudes of  $G$-TQFT2/$G$-CTST  and holographically dual to the two-dimensional theory.   The topological nature is reflected in the fact that we have not chosen a non-vanishing  Hamiltonian to define a time evolution, and all the interesting amplitudes, probabilities
and their inter-relations are encoded in the overlaps involving the interesting elements of $\cH$ such as the conjugacy class observables $ T_p$, the projectors $P_R$ and the handle creation operator $\Pi$.  The baby-universe creation operators \eqref{BabyUC} are operators  on   $S^{\infty} ( \cH) $. 

\subsection{Probabilities in $G$-CTST } 

In $G$-CTST it is natural to consider the partition function that results if we sum over all possible $h$ values.
The resulting partition function continues to have an interpretation as an ensemble average.
If we sum over genera, then the norm of the ground state is
\bea
\langle 0|0\rangle =\sum_{h=0}^\infty Z_{\Sigma_h}g_{st}^{ 2h-2}
=\sum_{h=0}^\infty \sum_R \left({|G|^2\over d_R^2}\right)^{h-1}g_{st}^{ 2h-2}
\label{WCoupling}
\eea
where $g_{st}$ is the string coupling.
There is a sum over $h$ for each $R$, each of which looks like a geometric progression, with radius
\bea
r={|G|^2 g_{st}^2\over d_R^2}
\eea
The sum over $h$ converges as long as
\bea
g_{st}^2 < {d_R^2\over |G|^2}
\eea
Performing the geometric sum, we find
\bea
\langle 0|0\rangle =\sum_R {{d_R^2\over |G|^2 g_{st}^2}\over 1-{|G|^2 g_{st}^2\over d_R^2}}
=\sum_R {d_R^4\over g_{st}^2|G|^2(d_R^2 -|G|^2 g_{st}^2)}
\eea
Notice that the condition ensuring that the geometric sum converges also ensures that the norm of $\langle 0|0\rangle$ is
positive.

The partition function is
\bea
Z_{\Sigma_{T_{p_1}\cdots T_{p_n }T_{q_1}^*\cdots T_{q_m }^*}}
&=&\langle \hat{T}^\dagger_{q_1}\cdots \hat{T}^\dagger_{q_m }\hat{T}_{p_1}\cdots \hat{T}_{p_n}\rangle\cr
&=&\sum_R\sum_{h=0}^\infty 
\left({|G|^2\over d_R^2}\right)^{h-1}{\chi_R(T_{p_1})\over d_R}\cdots {\chi_R(T_{p_n })\over d_R}
{\chi_R(T_{q_1}^*)\over d_R}\cdots {\chi_R(T_{q_m }^*)\over d_R}\cr
&\equiv& \mathfrak{z}\sum_R\sum_{h=0}^\infty p_R {\chi_R(T_{p_1})\over d_R}\cdots {\chi_R(T_{p_n })\over d_R}
{\chi_R(T_{q_1}^*)\over d_R}\cdots {\chi_R(T_{q_m }^*)\over d_R}
\eea
where 
\bea
p_R=\left(\sum_T{d_T^4\over g_{st}^2|G|^2(d_T^2 -|G|^2 g_{st}^2)}\right)^{-1}
{d_R^4\over g_{st}^2|G|^2(d_R^2 -|G|^2 g_{st}^2)}
\eea
It is obvious that
\bea
\sum_R p_R = 1
\eea
Again, we can write the partition function as
\bea
{Z_{\Sigma_{T_{p_1}\cdots T_{p_k}T_{q_1}^*\cdots T_{q_l}^*}}\over \langle 0|0\rangle }
=\sum_R p_R \langle 0|\hat{T}^\dagger_{q_1}\cdots \hat{T}^\dagger_{q_l}\hat{T}_{p_1}\cdots \hat{T}_{p_k}|0\rangle
\eea
which illustrates the fact that we can interpret the partition function as computing observables in a classical 
ensemble of theories. As in section \ref{PlanchTwo} the amplitudes and probabilities considered and their inter-relations can expressed within a topological quantum mechanics based on $\cH$ and $ S^{ \infty} ( \cH) $.

\section{S-duality for  $G$-CTST } \label{Sdual}

The calculations performed  in sections \ref{sec:Construct} and \ref{sec:characters}  have established that $G$-CTST provides a construction of the dimensions $d_R$ and the characters
of the irreducible representations of any finite group.
These constructions are examples of Fourier transforms, valid for general groups, between representation theory 
data and group-combinatoric data. 
In this section we will show that the same methods can be used to give an interpretation for sums of positive powers of
the dimensions of irreducible representations, including for example
\bea 
\sum_{R} d_R^4 \label{positivepower}
\eea
in terms of $G$-TQFT2 with defects, for any finite group $G$. 
Further, we will argue that this construction is intimately related to the S-duality transformation of $G$-CTST.

The group algebra $ \mC( G ) $ has an  inner product
\bea
\langle g_1|g_2\rangle = \delta (g_1g_2^{-1})
\eea
Consider the algebra
\bea 
\mC ( G ) \otimes \mC ( G ) 
\eea
There is an operator $X$  acting on the algebra as a projector, whose image has dimension $ \sum_R d_R^4$.
The projector $X$ is defined by
\bea 
X = \sum_{ R } P_R \otimes P_R\qquad
P_R={d_R\over |G|}\sum_{ g \in G}\chi_R( g ) g^{ -1} 
\eea
Indeed, a straight forward computation gives
\bea 
\Tr (X)&=&\sum_{g_1,g_2\in G}\langle g_1,g _2|X|g_1,g_2\rangle\cr
&=&\sum_R\sum_{g_1,g_2}\delta (g_1^{-1}P_Rg_1)\delta (g_2^{-1}P_Rg_2)\cr
&=&\sum_R |G|^2 \delta(P_R)\delta(P_R)\cr
&=&\sum_R d_R^2\sum_{g_1\in G}\chi_R(g_1)\delta(g_1^{-1} )
\sum_{g_2\in G}\chi_R(g_2)\delta(g_2^{-1} )\cr
&=&\sum_R d_R^4 
\eea
providing a construction of the sum (\ref{positivepower}).
This is a partition function on a product of two tori, each of which has a single boundary.
To see this, note that the geometrical interpretation of the delta functions is as follows 
\bea
\begin{gathered}\includegraphics[scale=0.4]{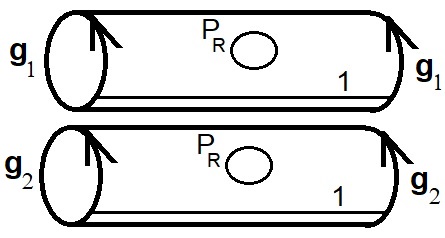}\end{gathered}
=\delta(g_1^{ -1 } P_R g_1)\delta(g_2^{-1} P_R g_2)
\eea
Summing over $g_1$ and $g_2$ closes the cylinders into tori
\begin{align*}
\sum_R \begin{gathered}\includegraphics[scale=0.4]{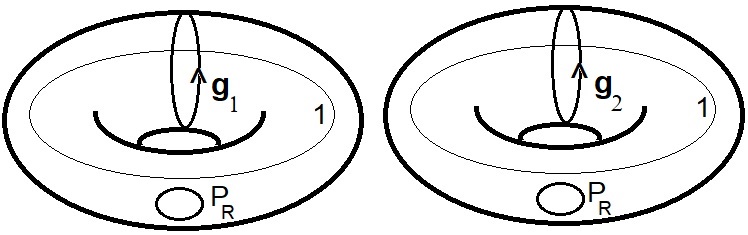}\end{gathered}
=\sum_R \sum_{g_1}\delta ( g_1^{ -1 } P_R g_1 ) 
\sum_{g_2}\delta ( g_2^{-1} P_R g_2 )=\sum_R d_R^4
\end{align*}

A simple extension of the logic above can be used to give a formula for $\sum_{R}d_R^{2k}$ for any positive
integer $k$.
In this case the operator playing the role of $X$ is given by
\bea
\sum_R P_R\otimes P_R\otimes \cdots\otimes P_R
\eea
where there are $k$ factors in the tensor product. 

There is an alternative description for $X$, which does not explicitly use the projectors $P_R$. We will give a description of this for the case where $G$ is symmetric group $S_n$, where known facts about the centre of $\cZ( \mC( S_n))$ allow a concrete discussion.  We will make use of $T_k\in \cZ( \mC(S_n)) $ given by summing all permutations with a single non-trivial cycle of length $k$.
This alternative description follows by exploiting the fact that conjugacy classes labelled by partitions of $n$ provide a basis
for the centre of the group algebra $\mC(S_n)$, denoted as $\cZ(\mC(S_n))$\cite{KR1911}. 
It turns out that a subset of these basis elements, those given by $T_k$ with $k\le k_* (n)$, will generate 
$\cZ(\mC(S_n))$\cite{KR1911}.
$k_*(n)$ is a (not explicitly known) function of $n$ whose form is determined by the degeneracies in the characters of $S_n$.
The projectors $P_R$ associated with irreducible representations of $S_n$ also generate $\cZ(\mC(S_n))$, so it is not
surprising that these two possibilities exist.
The alternative formula for $X$ follows by noting that the null space of $H$
\bea 
  H=\sum_{k=1 }^{ k_*(n) }   ( T_k \otimes  1 - 1 \otimes T_k ) 
\eea
is the image of $X$.
To see why this is the case, note that $T_k$ is in $\cZ(\mC(S_n))$, so that any state belonging to an irreducible
representation $R$ is an eigenstate of $T_k$ with eigenvalue determined by $R$.
If two states have the same eigenvalue for each $T_k$ with $k\le k_* (n)$, they must belong to the same irrep.
Thus, states in the null space of $H$ are sums of tensor products of states, where each term tensors pairs of states
that belong to the same irrep.
This is clearly equal to the image of $X$. In order to apply  this construction of $X$ to general $G$-TQFT2, we need to  develop the results analogous to those of \cite{KR1911} for other $G$, i.e. identify sets of generating conjugacy class sums for the centre $ \cZ( \mC (G))$. These generating sets can be chosen to be appropriate sets of conjugacy bbc news classes with small sizes. 

This completes the discussion of the construction problem for positive power sums of $d_R$. We now  explain the relevance of these constructions to the S-dual of $G$-CTST. 
Consider the sum of genus $h$ $G$-TQFT2 partition functions weighted by powers of the string coupling, which defines the partition function $Z (g_{st})$ of  $G$-CTST
\bea 
&& Z ( g_{st} ) = \sum_{h =0}^{ \infty }
g_{st}^{ 2h-2} Z_{ \Sigma_h }  =  \sum_{h =0}^{ \infty }
g_{st}^{ 2h-2} \sum_{R} \left ( { |G| \over d_R} \right )^{ 2h - 2 } \cr 
&& = \sum_{R} { d_R^2 \over |G|^2 g_{st}^2 }  { 1 \over ( 1 - g_{st}^2 |G|^2/d_R^2  ) }\eea
Defining the dual string coupling  $\tilde{g}_{st}=g_{st}^{-1}$  we have
\bea
Z (\tilde g_{st})
=\sum_R {d_R^4\tilde{g}_{st}^4\over |G|^2(\tilde{g}_{st}^2d_R^2 -|G|^2)}
\eea
Studying this expression for small values of the dual coupling, defines a new $S$-dual  expansion
\bea
Z (\tilde  g_{st})=-\sum_R\sum_{n=0}^\infty\tilde{g}_{st}^{2+2n} {d_R^{4+2n}\over |G|^{4+2n}}
\equiv\sum_{h}c_h \tilde{g}_{st}^{2h-2}
\eea
We see immediately that the dual partition function is written in terms of positive power sums of $d_R$ and that
the genus of the surfaces being summed is
\bea
h=2+2n
\eea
For example, the term with $n=0$ is precisely the product of two tori described above.

\subsection{Singularities of the partition function $G$-CTST as a function of string coupling  }\label{sec:singularities} 

 It is instructive to consider the  analytic structure of $ Z ( g_{st}^2 ) $  in the general  complex $g_{st}^2$ plane and identify how group theoretic data appears in the singularity structure, i.e. the poles and residues of $ Z ( g_{st} ) $. To describe the simplest connections, it is convenient to consider the sum over genuses from  $h=1$ to infinity, which we denoted as $Z_{1^+} (g_{st}^2)$
We will find that  this partition function has poles  at $ g_{st}^2 =a_R^{-2}$ with 
$a_R\equiv{|G|\over d_R}$. 
It is possible for two distinct irreps to have the same dimension, so that we can have $a_R=a_{R'}$, even when $R\ne R'$.
Denote the distinct values of $a_R^{-2}$ by $v_i$.
The multiplicity $m_{v_i}$, of $v_i$ is given by the residue of the pole at $v_i$. 
These assertions are easily established with a simple computation
\bea 
Z_{ 1^+} ( g_{st}^2 ) &=& \sum_{h=1}^\infty g_s^{2h-2} Z_{ h} 
= \sum_{ h=1 }^{ \infty }  \sum_{ R } \left({|G|\over d_R}\right)^{  2h  - 2 } g_s^{ 2h-2} \cr 
&& = \sum_{R}  { 1 \over ( 1 - a_R^{2} g_s^{2} ) }  \cr 
&& = \sum_{R}   { a_R^{-2}  \over ( a_R^{-2} - g_s^{  2} ) }  \cr 
&& = \sum_{i=1}^{i_*} { - v_i^2 m_{v_i}  \over ( g_s^2 - v_i^2 ) } \label{pertZ}
\eea
where there are a total of $i_*$ distinct $v_i$ values.
Given the exact analytic form of the partition function, it is possible to generate a number of expansions, valid for different
values of the coupling.
The expansion shown on the first line of (\ref{pertZ}) above is for weak coupling, when the string coupling is smaller than all 
of the $v_i$.
Notice that dimensions of irreps are raised to negative powers.
The strong coupling expansion is valid when the string coupling is larger than all of the $v_i$.
In this case, the expansion is
\bea 
Z_{ 1^+} ( g_{st}^2 )  &=& -\sum_{i=1}^{i_*} {v_i^2 m_{v_i}  \over ( g_s^2 - v_i^2 ) }\cr
&=& -\sum_{i=1}^{i_*} {v_i^2 m_{v_i}\over g_s^2}\sum_{n=0}^\infty \left({v_i^2\over g_s^2}\right)^n\cr
&& = -\sum_R { d_R^2 \over |G|^2 g_s^2}\sum_{n=0}^\infty \left({d_R^2\over |G|^2 g_s^2}\right)^n
 \label{strongcZ}
\eea
Notice that dimensions of irreps are now raised to positive powers.

There are other possibilities that generalize the weak coupling and strong coupling expansions.
Choose the index $i$ so that the $v_i^2$ are ordered, i.e. $v_1^2<v_2^2<\cdots<v_{i_*}^2$.
The more general expansions we could consider are obtained by choosing a value of the string coupling $g_s^2$ 
which is greater than $v_1^2$ but smaller than $v_{i_*}^2$.
These more general expansions can not be written as power expansions in either $g_s^2$ or its inverse, but rather
they are Laurent expansions in $g_s^2$.
To develop these expansions, introduce the two sets $i_<$ and $i_>$ defined by
\bea
i_<=\{ i|v_i^2 < g_s^2\}\qquad
i_>=\{ i|v_i^2 > g_s^2\}
\eea 
The most general expansion of the partition function can now be written as
\bea
Z_{ 1^+} ( g_{st}^2)=\sum_{h=1}^\infty\sum_{i\in i_<}m_{v_i}v_i^{2-2h}g_s^{2h-2}
-\sum_{h=1}^\infty\sum_{i\in i_>}m_{v_i}v_i^{2+2h}g_s^{-2-2h}
\eea
Note that when $i_<$ is empty we reproduce the strong coupling expansion of Section \ref{Sdual}
and when $i_>$ is empty we reproduce the weak coupling expansion.

\section{ Finiteness  relations in $G$-TQFT2 }\label{sec:finiteK} 

In this section, we describe relations between  amplitudes in $G$-TQFT2/$G$-CTST due to the finiteness of $G$. The relations we describe are focused on amplitudes for closed surfaces and surfaces with boundary. They depend on the dimension of the centre $ \cZ ( \mC( G) )$ which we denote as $K$. We have seen the key role played by matrices of size $K$ in the Sections \ref{sec:Construct} and \ref{sec:characters} on the construction of representation theoretic data from group multiplication combinatorics shaped by the amplitudes. The algebra $ \cZ ( \mC( G) )$  has also been prominent in the description of probability distributions associated with $G$-TQFT2 and $G$-CTST in section \ref{sec:probabilities}. In this section, we describe universal finite $K$ relations. We draw on  some mathematical analogies between  $G$-TQFT2 and the BPS sectors of $AdS_5/CFT_4$, based on the fact that the partition functions of $G$-TQFT2 are expressible in terms of traces of powers of a matrix. This allows us to define a simple inner product, such that the finite $K$ relations appear as null states in the inner product. The inner product contains a large $K$ factorization which fails when $1/K$ corrections are taken into account. The failure of factorization has a geometrical interpretation in terms of a mixing between  surfaces with different numbers of   connected components. We give a 2D topological field theory formulation of the inner product by coupling $G$-TQFT2 to TQFT2 based on symmetric group algebras. We describe this coupled toplogical field theory as $ \mC( G ) \times (\mC( S))_{ \infty }$ TQFT2.

\subsection{  Finite $K $ relations from null states of an inner product }\label{sec:Kinnerproduct}

Pick a finite group $G$ and let $K$ be the number of conjugacy classes. It determines a matrix $X = {\Diag} ( |G|^2  / d_R^2  ) $ of size $K$.  $G-TQFT2$  associates to a surface of genus $h$ the partition function 
\bea 
Z_{ \Sigma_h }   = \tr X^{ h -1}  \equiv Z_h  
\eea
A  disconnected surface has a partition function which is a product over the connected components, e.g. with two components we 
\bea 
Z  ( \Sigma_{ h_1} \times \Sigma_{ h_2} ) = \tr X^{ h_1 -1}  \tr X^{ h_2 -1}  
\eea
This can be computed by lattice TQFT2 on the two surfaces \cite{Witten2dYM,FHK}. 

The finite $K$ relations ensure that higher genus connected partition functions 
can be written in terms of linear combinations of partition functions for disconnected surfaces. 
This is a consequence of the Cayley Hamilton relations, which state that any matrix $X$ obeys its own eigenvalue equation. We have, using the elementary symmetric polynomials  $e_k(X)$ described in section \ref{sec:Construct}   
\bea 
X^K + \sum_{ k =1}^{ K } (-1)^k  e_k ( X ) X^{ K - k }  =0 
\eea
Multiply this with $ X^{l}$ for any positive integer $l$ 
\bea 
X^{ K + l  } + \sum_{ k =1}^{ K } (-1)^k  e_k ( X ) X^{ K - k + l   }  =0 
\eea
Taking  a trace on both sides, we have  
\bea 
\tr ( X^{ K + l } ) + \sum_{ k =1}^{ K } (-1)^k  e_k ( X ) \tr ( X^{ K - k + l  } )   =0 \, . 
\eea
For $l=1$, this gives $\tr X^{K+1}$ as a linear combination of products of traces of lower powers of $X$ than the $K$'th power.   Writing the elementary symmetric functions $e_{k}(X)$ in terms of traces of $X$ using \eqref{ekformula}, we 
obtain the trace relation
\bea\label{trRel}  
\tr X^{ K +1 } = \sum_{ k =1}^{ K } \tr X^{ K - k + 1 } \sum_{ p \vdash k }  { (-1)^{ 1 + \sum_{ i } p_i } \over \prod_{ i } i^{ p_i} p_i! }  \prod_{ i } ( \tr X^i )^{ p_i}  
\eea
For example if $ K =3$ we have 
\bea 
\tr  X^4 = \tr  X \tr X^{3  }   - { 1 \over 2 } ( \tr X^2 + ( \tr X )^2 ) \tr X^{ 2 } + 
{ 1 \over 6 } ( 2 \tr X^3 +  3  \tr X^2 \tr X + ( \tr X)^3       )   \tr X 
\eea
The relation \eqref{trRel} implies that the genus $K+2$ partition function $Z_{K+2}=\tr(X^{K+1})$ can be expressed in 
terms of products of smaller genus partition functions
\bea 
Z_{ K +2 } = \sum_{ k =1}^{ K } Z_{ K - k +2 }    \sum_{ p \vdash k  }  { ( -1)^{ 1  + \sum_{ i } p_i }  \over \prod_{ i } i^{ p_i} p_i! }
 \prod_i ( Z_{ i +1}  )^{ p_i }
\eea
To obtain this result we have used the expression for the partition functions in terms of traces. 

These equations raise the  interesting question of how to describe the finite $K$ relations  among $G$-TQFT2 amplitudes in generality. The  matrix form of the partition function $ Z_h = \tr ( X^{ h -1} ) $  relates this question to 
the description of finite $N$ effects in the AdS/CFT correspondence \cite{malda,witten,gkp}, 
specifically the half-BPS sector of string theory in $ AdS_5 \times S^5$ which corresponds to $ \cN =4$ super-Yang-Mills theory with $U(N)$ gauge group. Such finite $N$ effects are associated with the very rich physics of the stringy exclusion principle and giant gravitons \cite{malstrom,mst}. The half-BPS states of $ \cN=4$ SYM  for $U(N)$ gauge group correspond (by the operator-state correspondence of CFT) to  multi-traces of a complex matrix $ Z$. 
An orthogonal  basis for these states using the free field inner product 
in the $U(N)$ theory is labelled by Young diagrams \cite{CJR}.  If we consider gauge invariant states (multi-traces)  of dimension $n$, i.e. containing $n$ copies of $Z$, in the range $ n > N$, there is subspace of the vector space of these trace operators which vanishes due to finite $N$ relations (Cayley-Hamilton relations discussed above). 
A linear basis for these vanishing states is labelled by Young diagrams $R$ with first column of length greater than $N$. These are null states for the $N$-dependent inner product for multi-trace structures which comes from free field 
$U(N)$ theory. 

In order to answer the question of a systematic description of finite $K$ relations  for a group $G$ with $K$ conjugacy classes, using 
the  above  technical perspectives from the mathematics  of the stringy exclusion principle in AdS/CFT, it is convenient to introduce an abstract polynomial algebra $ \cP ( \omega_1 , \omega_2 , \cdots )$. This is a vector space over $ \mC$. $\omega_k$ corresponds to a  topological surface of genus $k+1$. We have a generator $ \omega_1 $ for a surface of genus $2$, a generator $ \omega_2$ for a surface of genus $3$ etc. A monomial $ \omega_1^{ k_1} \omega_2^{ k_2} \cdots  \omega_{ l }^{ k_l }$ corresponds to a disjoint union of $k_1$ genus two surfaces, $ k_2$ genus $3$ surfaces, up to $ k_l$ genus $l+1$ surfaces.  These finite monomials form a basis for the vector space $ \cP ( \omega_1 , \omega_2 , \cdots )$. The vector space $ \cP ( \omega_i)$ is endowed with a product defined 
as the product of these monomials. $G$-TQFT2 associates to the monomial 
\bea\label{GTFTmapo}  
&& \hbox{ $G$-TQFT2 } : \omega_1^{k_1 } \omega_2^{ k_2} \cdots \omega_{ l }^{ k_l } \rightarrow \Sigma_{2}^{ \times k_1 }\times \Sigma_3^{ \times k_2 }\cdots \Sigma_{ l+1 }^{ k_l } 
\rightarrow ( \tr X )^{ k_1}  ( \tr X^{ 2} )^{ k_2} \cdots ( \tr X^{ l } )^{ k_l }   \cr 
&& 
\eea
where $X$ is a $K \times K$ diagonal matrix, with entries labelled by irreps $R$  of $G$, and taking 
values $ { |G|\over d_R } $. 
$AdS_5/CFT_4$ associates to the same monomials half-BPS states corresponding to matrix traces 
\bea 
\hbox{AdS$_5$/CFT$_4$ } : \omega_1^{k_1 } \omega_2^{ k_2} \cdots \omega_{ l }^{ k_l } 
\rightarrow ( \tr Z )^{ k_1}  ( \tr Z^{ 2} )^{ k_2} \cdots ( \tr Z^{ l } )^{ k_l }   
\eea
where $Z = X_1 + i X_2$ and $X_1 , X_2$ are two of the $6$ hermitian matrices of $U(N)$ $\cN=4$ SYM. 
The parameter $N$ in the AdS5/CFT4 context is analogous to $K$ in the case of $G$-TQFT2.

The computation of 2-point functions of general holomorphic trace of dimension $n$  in $ \cN=4$ SYM 
with another general anti-homolomorphic trace defines an inner product on the space of traces. The outcome of the computation can be expressed  using permutations in the symmetric group $S_n$. Consider the 2-point function  
\bea 
\langle ( \tr Z )^{p_1}  ( \tr Z^2 )^{ p_2} \cdots ( \tr  Z^{ l }  )^{p_l } 
  ( \tr Z^{ \dagger}  )^{q_1}  ( \tr ( Z^{ \dagger} )^2 )^{ q_2} \cdots ( \tr  (Z^{ \dagger})^{ l }  )^{q_m }   \rangle 
\eea
where the holomorphic operator has a trace structure specified by
 the exponents $(p_1, p_2,\cdots p_l)\equiv p$,
  and the anti-holomorphic  operator has a trace structure
 specified by $(q_1,q_2, \cdots ,q_m)\equiv q$. 
 For fixed dimension $n$, we have $ n = \sum_{ i  } i p_i = \sum_{ i } i q_i$. $p$ and $q$ are partitions of $n$, which also correspond to conjugacy classes of $S_n$. The 2-point function 
determines the inner product \cite{CJR} 
 \bea\label{HBPSIP}  
 && \langle ( \tr Z )^{p_1}  ( \tr Z^2 )^{ p_2} \cdots ( \tr  Z^{ l }  )^{p_l } 
  ( \tr Z^{ \dagger}  )^{q_1}  ( \tr ( Z^{ \dagger} )^2 )^{ q_2} \cdots ( \tr  (Z^{ \dagger})^{ l }  )^{q_m }   \rangle  \cr 
  &&  = { n! \over |\cC_p| |\cC_q|}   \sum_{ \sigma_1 \in \cC_p  } \sum_{ \sigma_2 \in \cC_q } \sum_{ \sigma_3 \in S_n } 
\delta ( \sigma_1 \sigma_2 \sigma_3 ) N^{ C_{ \sigma_3} } 
 \eea
 where $ C_{ \sigma_3 } $ is the umber of cycles in $ \sigma_3$. 
 The Schur-basis of gauge invariant operators are labelled by Young diagrams $Y$ with $n$ boxes
 \bea 
 \chi^Y ( Z ) = { 1 \over n!  } \sum_{ \sigma \in  S_n } \chi^Y ( \sigma ) \cO_{ \sigma} ( Z )  
 \eea
 where 
\bea 
\cO_{ \sigma } ( Z ) = Z^{ i_1}_{ i_{ \sigma (1) }  }  \cdots Z^{ i_n }_{ \sigma (n ) } 
\eea 
The two point function in the Schur basis is given by
 \bea 
 \langle \chi^{ Y_1 } ( Z ) \chi^{ Y_2} ( Z^{ \dagger} ) \rangle
 = \delta^{ Y_1, Y_2 } f_{ Y_1  } 
 \eea
 The normalization factor $f_Y$  for a Young diagram  is a polynomial in $N$ equal to 
 \bea 
 f_Y = \prod_{(i , j) } ( N - i + j )
 \eea
 where $(i,j)$ are row and column labels of the boxes in the Young diagram $Y$. The norm of all Young diagram states with $Y$ having more than $N$ rows vanishes and in fact the polynomials $ \chi^Y ( Z ) $ are identically zero for $Z$ of size $N$. Thus the inner product \eqref{HBPSIP} for trace structures encodes finite $N$ relations on traces  in the form of null states for an inner product. 
  
We  give  the algebra $ \cP ( \omega_i ) $ an inner product depending on a parameter $K $ of the form familiar from the matrix combinatorics \eqref{HBPSIP} of the half-BPS sector in AdS5/CFT4: 
\bea\label{InnProdHalf}  
\langle \omega_1^{ p_1 } \omega_2^{ p_2} \cdots | \omega_1^{ q_1} \omega_2^{ q_2} \cdots \rangle 
= { n! \over |\cC_p | |\cC_q| }\sum_{ \sigma_1 \in \cC_p  } \sum_{ \sigma_2 \in \cC_q  } \sum_{ \sigma_3 \in S_n } 
\delta ( \sigma_1 \sigma_2 \sigma_3 ) K^{ C_{ \sigma_3} } 
\eea
With this inner product, all the finite $K $ relations are null states and can be expressed 
in terms of Schur Polynomials of $X$. Using the $G$-TQFT2 map, this corresponds to an inner product between surfaces  $ \prod_i \Sigma_{i+1}^{ \times p_i }  $ and 
$\prod_{ i } \Sigma_{ i+1}^{ q_i}$, with $ n = \sum_i i p_i = \sum_i i q_i$. 
The Schur basis for algebra $\cP ( \omega_i ) $ is obtained by replacing 
\bea 
\chi^Y (X) = { 1 \over n ! } \sum_{ p \vdash n }{ 1 \over |\cC_p| }  \chi^Y ( T_p) \prod_i \tr (X^i)^{ p_i} 
\rightarrow  \chi^Y ( \omega )={ 1 \over n ! } \sum_{ p \vdash n }  { 1 \over |\cC_p| }\chi^Y ( T_p ) \prod_{ i } ( \omega^i )^{ p_i}  
\eea
The inner product in the Schur basis is 
\bea 
\langle \chi^{Y_1}  ( \omega )  | \chi^{ Y_2}  ( \omega ) \rangle 
= f_{Y_1} \delta^{Y_1Y_2 } 
\eea

{\bf Factorization and $1/K$ corrections to factorization:} 
The inner product \eqref{InnProdHalf} factorizes in the leading  large $K$ limit. In this limit the dominant term comes from the case where $ \sigma_3$ has the maximum number of cycles, i.e. $ C_{ \sigma_3} = n $ and $\sigma_3$ is the identity permutation. In this case, $\cC_p = \cC_q$ and the inner product is non-zero only when $p$ and $q$ describe the same trace structure. At subleading orders in $ K$, other permutations $\sigma_3 $ contribute and the precise departures from $ p =q$ are encoded in permutation products. In the $G$-TQFT2 interpretation of the inner product, the factorization means that different monomials in $ \omega_i$,   which correspond to surfaces with different numbers of connected components, are orthogonal at large $K$ and there are corrections to this factorization at sub-leading orders in $1/K$. In the context of AdS5/CFT4, the departures from factorization formed an important argument in guiding the identification of CFT operators for  giant gravitons \cite{BBNS,CJR}.

The inner product we have considered is not the unique inner product that is compatible with the finite K relations. 
Other possible inner products are determined by Casimirs  as follows
\bea 
\langle \omega_1^{ p_1 } \omega_2^{ p_2} \cdots | \omega_1^{ q_1} \omega_2^{ q_2} \cdots \rangle 
= { n! \over |\cC_p | |\cC_q| } \sum_{ \sigma_1 \in \cC_p } \sum_{ \sigma_2 \in \cC_q } \sum_{ \sigma_3 \in S_n } 
\delta ( \sigma_1 \sigma_2 \sigma_3 C ) K^{ C_{ \sigma_3} } 
\eea
We have inserted a Casimir element  $C$ for $U(K)$ - expressed in terms of the group algebra of $S_n$. The existence of this map relies on Schur-Weyl duality and plays an important role in the string theory of 2D Yang Mills theory \cite{GrTa,CMR,Ganor} as well as AdS5/CFT4 \cite{EHS,KR1911}. 

This will still be diagonal in the Schur basis and would give a different normalisation of the 2-point function, 
modified by presence of the Casimir. The 2-point function in the Schur basis is now 
\bea 
\langle \chi^{ Y_1}  ( \omega ) | \chi^{Y_2}  ( \omega ) \rangle 
= f_{Y_1}  { \chi^{Y_1}  ( C ) \over d_{ Y_1} } 
\eea

\subsection{ Finiteness relations and $ \mC ( G  ) \times (\mC(S))_{ \infty} $  two dimensional topological field theory  }\label{sec:KSinfty} 

In the above discussion, we have found it useful to introduce a simple inner product for a polynomial algebra of 
Riemann surfaces, which captures the finite $K$ relations $G$-TQFT2.  The simplest inner product coincides with an inner product we have seen in the half-BPS sector of 
$ N=4$ SYM, but variations of the inner product which also capture the finite $K$ relations are also described. 
A natural question is: how do we interpret these inner products as a construction within Dijkgraaf-Witten theory?
Closely related to Dijkgraaf-Witten theory is the open-closed topological field theory developed by 
Moore and Segal \cite{MooreSegal}. In \cite{MooreSegal} the amplitudes of Dijkgraaf-Witten theory for closed surfaces and surfaces with boundary are interpreted in terms of the centre of the  associative algebra $\mC(G)$ which is equipped with a trace map which is the trace in the regular representation. There is also an extension to open strings which uses $ \mC ( G )$ and not just its centre, but we will not make extensive use of the open string sector in this paper.

Consider the central element $ \widehat Y_h  \in\cZ (\mC (G))$
\bea 
\widehat Y_h  = \Pi^{ h } 
\eea
where 
\bea\label{handlecreation} 
\Pi = \sum_{ g_1 , g_2 \in G } g_1 g_2 g_1^{-1} g_2^{-1}  = \sum_{ R } { |G |^2 \over d_R^2 } P_R 
\eea
As discussed in section \ref{sec:probabilities}
$\Pi$ can be viewed as a handle-creation operator. 
In terms of the handle creation operator, the partition function for genus $h$ is 
\bea\label{Zhdelt}  
Z_{ \Sigma_h } = { 1 \over |G |  } \delta ( \widehat Y_h ) 
\eea
This partition function also has an expansion,  in terms of Young diagrams, as 
\bea 
Z_{\Sigma_h } = \sum_{ R } \left  ( { |G | \over d_R } \right )^{ 2 ( h -1) }  = \tr X^{ h - 1 } 
\eea
where 
\bea  
X = { \Diag } \left ( { |G|^2 \over d_R^2 } \right  ) 
\eea
The handle creation operator $\widehat Y_h$  has an expansion in terms of central class elements $T_{\mu}$ as follows 
\bea\label{YhZ}  
\widehat Y_h  && = \sum_{  p  } \delta ( \widehat Y_h T_{ p}  ) { T_{ p }' \over { |\cC_{p }|  } } \cr 
&&  = \sum_{ p }\left (  { 1 \over |G| } \delta ( Y_h T_{ p  }  ) \right ) ~  T_{ p  }' ~ |{ \Sym~  \cC_{ p  } }| 
\eea
An important consequence of this expansion is that it can be used to develop an instructive geometrical interpretation.
The delta function is proportional to the partition function on a genus $h$ surface with a disc removed, 
$ \Sigma_{ h, 1}  $,  with 
\bea 
 Z_{ \Sigma_{ h, 1 }  ; T_p } = { 1 \over |G| } \delta ( \hat Y_h T_p ) 
\eea
The cylinder partition function defines an inner product 
\bea 
Z_{ cyl } ( T_p , T_q ) = { 1 \over |G |} \delta ( T_p T_q ) = \delta_{ \cC_p  , \cC_{ q'} } { | \Sym ( \cC_p ) |  \over |G|  }  =\delta_{ \cC_p  , \cC_{ q'} } { 1 \over |\cC_p | } 
\eea
Given the expansion (\ref{YhZ}), we can interpret $ \widehat Y_h$,  as a state in the Hilbert space $ \cZ ( \mC ( G ) ) $ 
corresponding to the genus $h$ surface with one boundary. 
Recalling that 
\bea 
{1 \over |G | } \delta ( \cdot ) 
\eea
is the function on $\cZ( \mC ( G ) ) $ given by disc partition function, we can interpret \eqref{Zhdelt} as the gluing of the 
disc to the genus $h$ surface minus a hole, as illustrated below
\bea
&&\qquad \Pi =\sum_{g_1,g_2\in G}g_1g_2g_1^{-1}g_2^{-1}\cr
\begin{gathered}
\includegraphics[width=0.3\columnwidth]{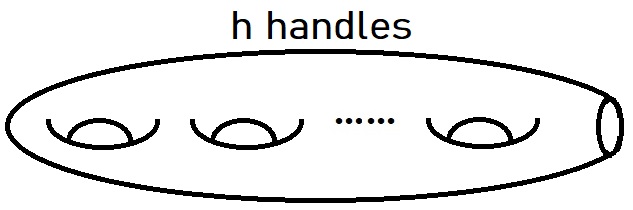}
\end{gathered}
\qquad&\rightarrow&\qquad\widehat{Y}_h=\Pi^h\in\cZ(\mC(G ))\cr
\left(\cZ(\mC(G ))\right)^*\ni {1\over |G|}\delta (\cdot)\qquad &\leftarrow&\qquad
\begin{gathered}
\includegraphics[width=0.035\columnwidth]{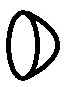}
\end{gathered}\cr
\begin{gathered}
\includegraphics[width=0.3\columnwidth]{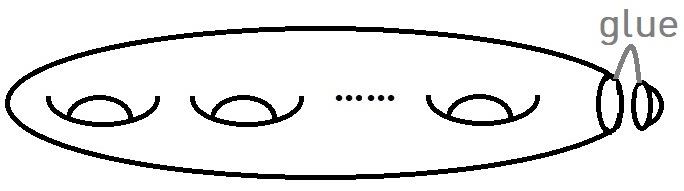}
\end{gathered}
\qquad&\rightarrow&\qquad
\begin{gathered}
\includegraphics[width=0.3\columnwidth]{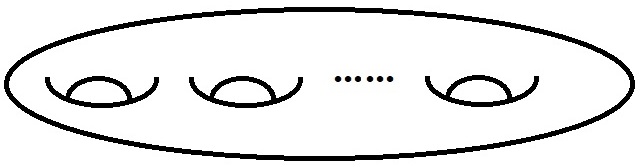}
\end{gathered}\cr
{1\over |G|}\delta(\widehat{Y}_h)\qquad &=&\qquad Z(\Sigma_h)
\eea

Developing the above discussion for a disconnected surface $\Sigma_{h_1}\times\Sigma_{h_2}\times\cdots\Sigma_{h_L}$ 
we find 
\bea 
 Z_{\Sigma_{h_1}\times\Sigma_{h_2}\times\cdots\times\Sigma_{h_L}} 
&=&Z_{\Sigma_{h_1}}Z_{\Sigma_{h_2}}\cdots Z_{\Sigma_{h_L}}\cr 
&=&\left ({1\over |G|}\delta\otimes {1\over |G|}\delta\otimes\cdots\otimes {1 \over |G|}\delta \right)\left( \widehat Y_{h_1}
\otimes\cdots\otimes\widehat Y_{ h_L } \right)
\label{DHandle} 
\eea
This has the interpretation of gluing $k$ discs to $k$ surfaces of genera $h_1,h_2,\cdots ,h_k$, each 
with a disc removed. The cut-and-paste operation produces the partition function from 
an element of $(\cZ (\mC (G)))^{\otimes L}$ corresponding to an element of $(\cZ (\mC (G)))^{*\otimes L}$. 
Diagrammatically, \eqref{DHandle} can be represented as
\bea
\begin{array}{cccc}
\begin{gathered}
\includegraphics[width=0.03\columnwidth]{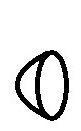}
\end{gathered}
&\begin{gathered}
\includegraphics[width=0.3\columnwidth]{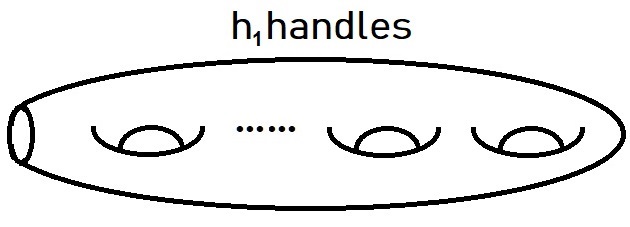}
\end{gathered}
&
&\begin{gathered}
\includegraphics[width=0.3\columnwidth]{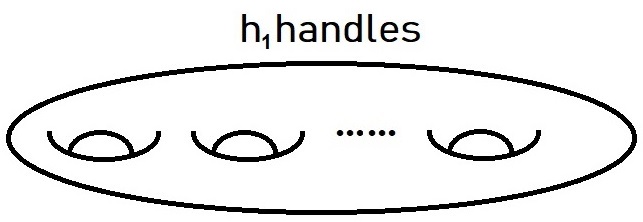}
\end{gathered}
\\
\begin{gathered}
\includegraphics[width=0.03\columnwidth]{hhandles}
\end{gathered}
&\begin{gathered}
\includegraphics[width=0.3\columnwidth]{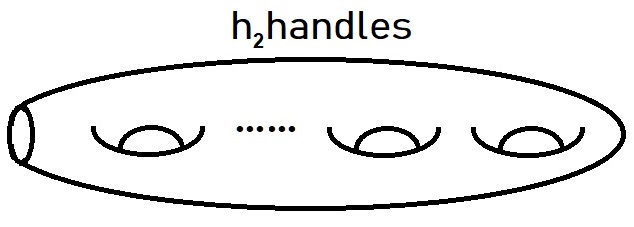}
\end{gathered}
&\longrightarrow
&\begin{gathered}
\includegraphics[width=0.3\columnwidth]{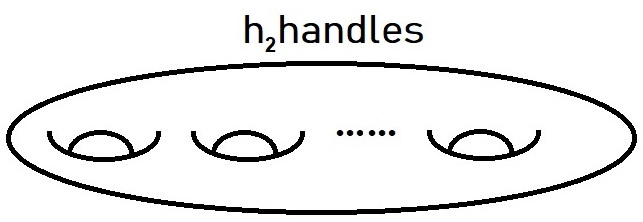}
\end{gathered}
\\
\vdots
&\vdots
&\vdots
&\vdots
\\
\begin{gathered}
\includegraphics[width=0.03\columnwidth]{hhandles}
\end{gathered}
&\begin{gathered}
\includegraphics[width=0.3\columnwidth]{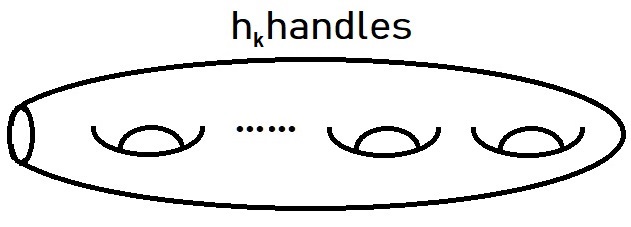}
\end{gathered}
&
&\begin{gathered}
\includegraphics[width=0.3\columnwidth]{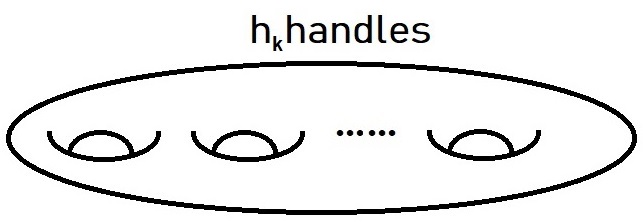}
\end{gathered}
\\
{1\over |G|}\delta\otimes {1\over |G|}\delta\cdots\otimes{1\over |G|}\delta
&\Pi^{h_1}\otimes\Pi^{h_2}\cdots\otimes\Pi^{h_k}
&\longrightarrow
&{1\over |G|^k}\delta(\Pi^{h_1})\cdots \delta(\Pi^{h_1})
\end{array}\nonumber
\eea

We have described an inner product for disconnected surfaces which accounts for the finite $K$ relations. This used symmetric groups $S_n$ for varying $n$, which is equal to the number of connected components in the surface being considered. Note that the formula for the inner product is itself given in terms of a delta function on the group algebra $\mC( S_n)$, which is suggestive of a permutation-TQFT2 interpretation of the inner product. The link between the  combinatorics and correlators  of $U(N)$ gauge theories, as well as theories involving products of $U(N_a)$, with symmetric group TQFT2  has been studied systematically  in \cite{QuivCalc,YusukeTFT}. Building on these results,  we show here  how the inner product \eqref{InnProdHalf} for $G$-TQFT2 amplitudes  can be given a geometrical interpretation by coupling the $ \cZ ( \mC ( G ) )$ theory to a theory based on the algebra 
\bea 
\bigoplus_{ n =0 }^{ \infty }( \mC ( S_n ) )   
\eea 
We will call this algebra 
$ ( \mC S )_{\infty } $. The construction we describe is based on TQFT2 for the algebra 
\bea 
 \cZ  ( \mC ( G ) ) \otimes (\mC S)_{ \infty }  
\eea
The first important ingredient in the construction of this theory is a cylinder which maps 
central elements in $\cZ( \mC( G) )$ to permutations $ \Pi^h \rightarrow ( 1, ... , h -1  ) $.

Consider the composition
\bea
\begin{gathered}
\includegraphics[width=0.3\columnwidth]{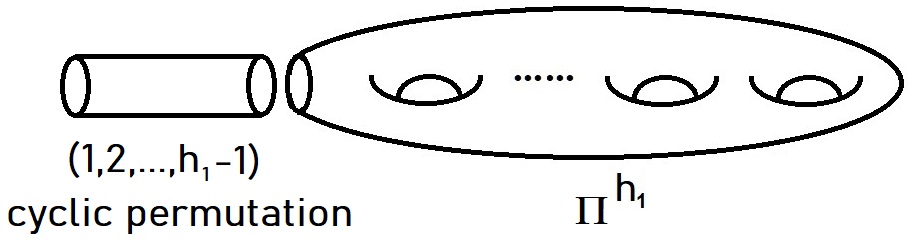}
\end{gathered}
\eea
The cylinder is a transition amplitude that takes in $\Pi^{h_1}$ and produces a permutation 
\bea
\begin{gathered}
\includegraphics[width=0.3\columnwidth]{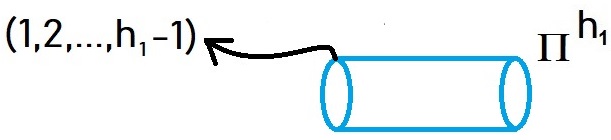}
\end{gathered}\label{cylndr}
\eea
This defines a transition between $\Pi^{h_1}\in\cZ(\mC(G))$ and a permutation.
Thus, the cylinder is something that is defined in $\cZ(\mC(G))\otimes(\mC S)_{\infty}$.

The formulae developed above can be used to given a diagrammatic interpretation to the inner product
\bea 
&&\langle\tr (X^{h_1-1})\tr (X^{h_2-1})\cdots\tr (X^{h_k-1})\tr (X^{h_1'-1})\tr (X^{h_2'-1})\cdots\tr (X^{h_l'-1})\rangle\cr 
&&=
{ n! \over |C_{ h_1 , \cdots , h_k } | |C_{ h_1' , \cdots , h_l' } |  }
\sum_{\sigma_1\in C_{h_1h_2\cdots h_k}}\sum_{\sigma_2\in C_{h_1'h_2'\cdots h_l'}}
\sum_{\sigma_3\in S_{\sum_i h_i}}
\delta (\sigma_1\sigma_2\sigma_3) K^{C_{\sigma_3}} \label{inrprd} \cr 
&& 
\eea
On the RHS above the notation $C_{a_1 a_2\cdots a_l}$ stands for the conjugacy class of permutations with cycle lengths
given by $a_1-1, a_2-1,\cdots,a_l-1$.
The above inner product is non-zero if and only if $\sum_i (h_i-1) =\sum_i (h_i'-1)$.
The partition function of the three holed sphere in $S_n$ TQFT2 of flat bundles, with boundary permutations $ \sigma_1 , \sigma_2 ,\sigma_3$   is 
\bea
\begin{gathered}
\includegraphics[width=0.2\columnwidth]{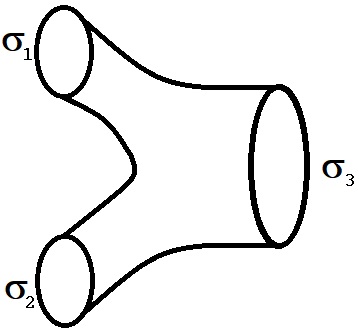}
\end{gathered}
=\sum_{\gamma_1,\gamma_2}
\delta (\gamma_1\sigma_1\gamma_1^{-1}\gamma_2\sigma_2\gamma_2^{-1}\sigma_3^{-1})
\eea
ensures that $\sigma_3$ lies in the product of the conjugacy class of $[\sigma_1]$ and the conjugacy class $[\sigma_2]$.
By introducing a unit defect as in \cite{QuivCalc} we have 
\bea
\begin{gathered}
\includegraphics[width=0.2\columnwidth]{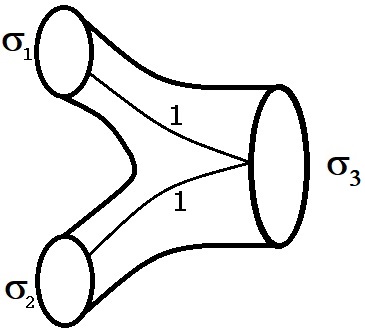}
\end{gathered}
=\delta (\sigma_1\sigma_2\sigma_3^{-1})
\eea
we set $\sigma_3=\sigma_1\sigma_2$.
The inner product (\ref{inrprd}) can now be expressed in terms of diagrams.
\begin{figure}[h]%
\begin{center}
\includegraphics[width=0.55\columnwidth]{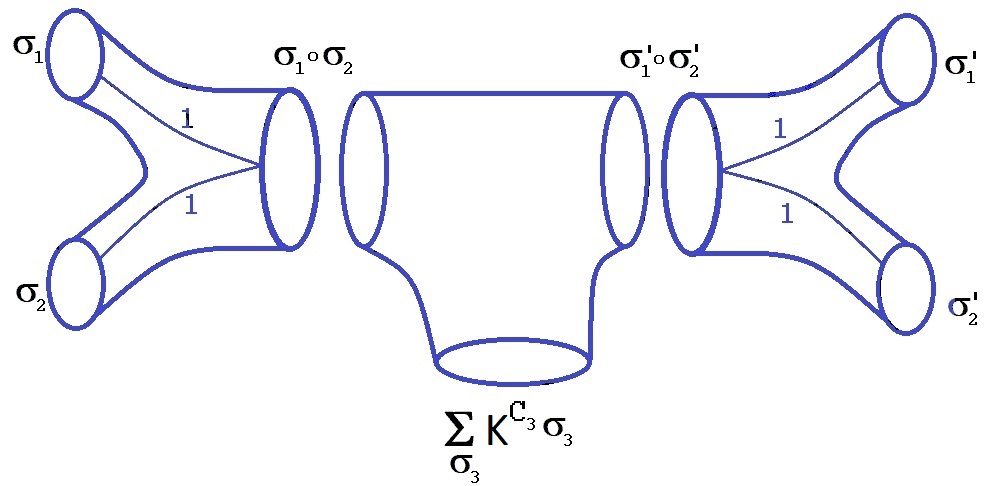}
\caption{The interpretation of the inner product (\ref{IPtwo}) in terms of diagrams. Here $\sigma_1=(1,2,\cdots,h_1-1)$,
$\sigma_2=(h_1,h_1+1,\cdots,h_2-2)$ and $\sigma_1\circ\sigma_2=(1,2,\cdots,h_1-1)(h_1,h_1+1,\cdots,h_2-2)$. 
There are similar formulas for the primed permutations.
The unit defect ensures that $\sigma_1\circ\sigma_2$ is a specific permutation and not a sum over a conjugacy class.
$C_3$ is the number of cycles in $\sigma_3$ and $\sigma\circ\psi$ is the outer products of permutations
$\sigma$ and $\psi$.} 
\label{fig:InnProd}
\end{center}
\end{figure}

\begin{figure}[h]\label{fig:fullinnerproduct}
\begin{center}
\includegraphics[width=0.95\columnwidth]{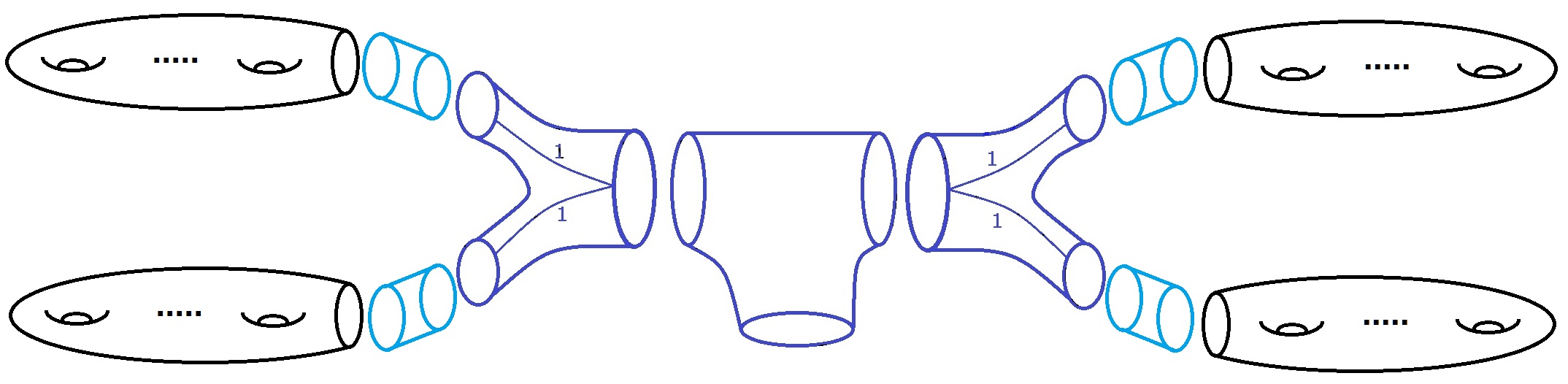}
\caption{The interpretation of the inner product (\ref{IPtwo}) in terms of diagrams.}
\label{Fig:InnProdOmeg}
\end{center}
\end{figure}

For clarity and because the generalization is immediate, consider the simpler formula
\bea 
\langle\tr (X^{h_1-1})\tr (X^{h_2-1})\tr (X^{h_1'-1})\tr (X^{h_2'-1}))\rangle=
{ n! \over |C_{ h_1 , h_2 }| | C_{ h_1', h_2'} |}
\sum_{\sigma_1\in C_{h_1h_2}}\sum_{\sigma_2\in C_{h_1'h_2'}}
\sum_{\sigma_3\in S_{\sum_i( h_{i}-1 )}}
\delta (\sigma_1\sigma_2\sigma_3) K^{C_{\sigma_3}}\cr
\label{IPtwo}
\eea
In terms of diagrams the formula \eqref{IPtwo} is displayed in Figure \ref{fig:InnProd}.
The inner product is non-vanishing if and only if $h_1+h_2=h_1'+h_2'$.
An  more complete picture incuding the coupling between the $G$-TQFT2 and the symmetric group sector ( using the cylinder \eqref{cylndr})
to write the inner product in terms of the TFT of the $(\mC S)_\infty\otimes \cZ (\mC (G))$ algebra. 
This provides an interpretation for the inner product as an amplitude on a 2-complex, as given in Figure \ref{Fig:InnProdOmeg}.  

\subsection{ Extensions of the finiteness discussion for closed surfaces } 

The discussion of the last subsection has related high genus surfaces without boundaries to
disjoint unions of lower genus surfaces, again without boundaries.
In this section we generalize this discussion first to surfaces with boundaries but fixed genus $h=1$, which involves the 
matrix $X_p$, and then to surfaces that have both multiple boundaries and any genus, which involves the pair $X,X_p$.

\subsubsection{ One-matrix finite $K$ relations for $ X_p$  }

By using the matrix $X$ above, we have described relations between high genus surfaces and
disjoint unions of lower genus surfaces. 
Similar relations hold for $X_p$, where $p$ is a conjugacy class. 
This will relate $ tr X_p^{ K + 1} $ with products of lower traces. 
Recalling \eqref{powersumXp} in Section \ref{sec:characters} the LHS is the partition function of a surface of one with $ K +1 $ holes 
each carrying the conjugacy class $p$. The RHS is for disjoint unions of genus one but with 
fewer boundaries.
The trace relation \eqref{trRel}, expressed in terms of the matrix $X_p$ of normalized characters 
for a conjugacy class $p$, is 
\bea\label{trRel}  
\tr X_p^{ K +1 } = \sum_{ k =1}^{ K } \tr X_p^{ K - k + 1 } \sum_{ p \vdash k }  { (-1)^{ 1 + \sum_{ i } p_i } \over \prod_{ i } i^{ p_i} p_i! }  \prod_{ i }  ( \tr X_p^i )^{ p_i}  
\eea
Since these traces are partition functions for surfaces with boundary 
conditions labelled by $p$, we have a relation  
\bea 
 Z (  \Sigma_{ h =1 ; \cC_p^{ \times ( K +1 )  } }   ) = 
\sum_{ k =1}^{ K }     Z (  \Sigma_{ h =1 ; \cC_p^{ \times ( K - k +1  )  } }   )
\sum_{ p \vdash k }  { (-1)^{ 1 + \sum_{ i } p_i } \over \prod_{ i } i^{ p_i} p_i! } 
 \prod_{ i } ( Z (  \Sigma_{ h =1 ; \cC_p^{ \times ( i  )  } }   ))^{ p_i} 
\eea

\subsubsection{2-Matrix finite $K$ relations  }

\begin{figure}[h]%
\begin{center}
\includegraphics[width=0.4\columnwidth]{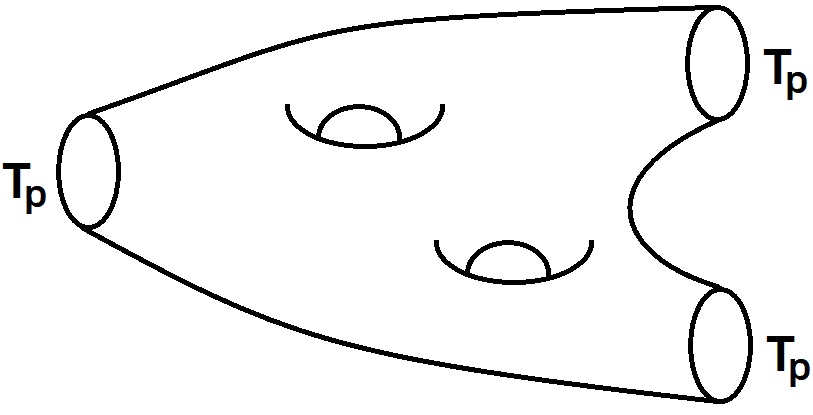}%
\caption{The surface shown has genus $h=2$ with $k=3$ boundaries.}
\label{Fig:2h3b}
\end{center}
\end{figure}

The partition function for a genus $h$ surface, with $k$ one-dimensional boundaries, each having the 
boundary condition that the holonomy around the boundary circle is in the conjugacy class $\cC_p$ is 
given by
\bea 
Z(\Sigma_{h;\cC_p,\cdots,\cC_p})=Z (\Sigma_{h;\cC_p^{\times k}}) = \tr ( X^{ h -1}  X_p^k )
\eea
where $X$ and $X_p$ are commuting diagonal matrices, with diagonal entries labelled by irreducible representations $R$ 
of $G$ 
\bea 
&& X = { \rm { Diag } }  \left ( { |G| \over d_R } \right ) \cr 
&& X_p = { \rm { Diag } } \left (   { \chi^R ( T_p ) \over d_R }  \right ) \label{explMats}
\eea
The finiteness of $K$  means that partition functions at 
high $h$ and high $k$ are  expressible in terms of products of traces of lower powers. 
For example consider
\bea 
Z(\Sigma_{h=K+1;\cC_p^{\times K+1}})=\tr ((X X_p)^{K+1}) 
\eea
The equation \eqref{trRel} with $X_p \rightarrow XX_p$ implies a finite $K$ relation 
relating this boundary partition function to lower boundary partition functions as follows
\bea
Z(\Sigma_{h=K+1;\cC_p^{\times (K+1)}}) = \sum_{k=1}^{K} Z(\Sigma_{h=K-k+1;\cC_p^{\times (K-k+1)}})
\sum_{p\vdash k}{(-1)^{1+\sum_{i}p_i}\over \prod_i i^{p_i}p_i!}\prod_i Z(\Sigma_{h=i;\cC_p^{\times i}})^{p_i}  
\eea

The systematic finite $K$ relations can be obtained using multi-symmetric functions. 
For ease of notation, we will write $ X_p  = Y $, so we are dealing with traces of two commuting matrices $X,Y$. 
There is a trace basis of functions of these two diagonal matrices, which is labelled by a sequence 
$[{\bf q}]\equiv [(q_{11},q_{12}),(q_{21},q_{22}),\cdots,(q_{M1},q_{M2})]$ with $q_{i1},q_{i2}\in\mathbb{N}_{0}$ 
where $\mathbb{N}_{0}$ is the set of natural numbers extended to include $0$: $(\mathbb{N})_0=\{0,1,2,\cdots\}$.
When
\bea 
&& \sum_i q_{1i} = q_1 \cr 
&& \sum_i q_{2i} = q_2 
\eea
$\bq$ is said to be a vector partition of $(q_1,q_2)$. 
In the first instance, it is useful to consider $K\gg M,q_{i1},q_{i2}$ so that all these sequences give a linearly independent
set of multi-symmetric functions
\bea 
 T_{[{\bf q}]}=\tr(X^{q_{11}}Y^{q_{12}})\tr(X^{q_{21}}Y^{q_{22}})\cdots\tr(X^{q_{M1}}Y^{q_{M2}}) 
\eea 
We will now introduce a second basis, which has an interpretation using coherent states in many-boson systems \cite{CLBSR}.
For our purposes, this second basis is particularly useful as it clarifies the origin of the finite $K$ relations.
Since $X$ and $Y$ commute, they can be simultaneously diagonalized.
Denote their eigenvalues as $x_i$ and $y_i$ $i=1,\cdots,K$ respectively.
In terms of these eigenvalues we motivate the second basis as follows. Consider 
\bea
T_{[(1,0),(0,1)]}&=&\Tr (X)\Tr (Y)=\sum_{i=1}^K\sum_{j=1}^K x_i y_j\cr
&=&\sum_{i\ne j=1}^K x_i y_j+\sum_{i=1}^K x_i y_i\cr
&=& M_{[(1,0),(0,1)]}+M_{[(1,1)]}\label{moti}
\eea
Notice that there are two sums in the first term above, and the indices for the two do not collide.
It is natural to interpret the first term above as a two particle state, with one of the particles having an ``$x$'' excitation 
and the second a ``$y$'' excitation.
The second term is a single particle state, which has both ``$x$'' and ``$y$'' excited.
See the original article \cite{CLBSR}, where this interpretation is developed in detail, for more background.
In general we have 
\bea
M_{[{\bf q}]}=\sum_{\sigma\in S_K}x_{\sigma(1)}^{q_{11}}y_{\sigma(1)}^{q_{21}}
x_{\sigma(2)}^{q_{12}}y_{\sigma(2)}^{q_{22}}\cdots x_{\sigma(M)}^{q_{1M}}y_{\sigma(M)}^{q_{2M}}
\eea
An important feature of this formula, is that there are now $M$ sums on the RHS and their indices never collide.
There is a linear transformation from the functions $T_{[{\bf q}]}$ to the polynomials $M_{[{\bf q}]} $ \cite{CLBSR}
\bea 
T_{[\bq]} = C_{\bq}^{~\br} M_{[\br]} 
\eea 
There is also an inverse transformation 
\bea 
M_{[\bq]}=\tilde C_{\bq}^{~\br} T_{[\br]} 
\eea
There is a connection between the matrices $C$, the inverse matrices $\tilde C$ and set partitions. 
The inversion of $C$ uses general theorems about set partitions that form a partially ordered set, as explained in 
Section 4.3 of \cite{CLBSR}. For the present discussion it suffices to  use the fact  that the inverse exists and the matrices $C,\tilde C$ are independent of $K$. 
This is  analogous to the fact that the transformation between Schur polynomials and traces in the 
1-matrix case is independent of matrix size and only depends of $n$, the degree of the traces being considered. 

To illustrate the above discussion we consider two examples.
For the first example, we consider the setting introduced in (\ref{moti}) above, corresponding to operators
$T_{[\bq]}$ constructed using a single $X$ and a single $Y$.
There are two possible vector partitions
\bea
[{\bf q_1}]=[(0, 1),(1, 0)]\qquad [{\bf q_2}]=[(1, 1)]
\eea
The matrices $C$ and $\tilde{C}$ are given by
\bea
C =\left[\begin{array}{cc}1 &1\\ 0 &1\end{array}\right]\qquad
\tilde{C}=\left[\begin{array}{cc}1& -1\\ 0&1\end{array}\right]
\eea
The matrix $C$ is easily read from (\ref{moti}) and it should be clear that $C$ is always upper triangular.
For a more interesting example, consider operators constructed using two $X$'s and a single $Y$.
In this case there are a total of four possible vector partitions
\bea
[{\bf q_1}]&=&[(0, 1),(1, 0),(1, 0)]\cr
[{\bf q_2}]&=&[(0, 1), (2, 0)]\cr
[{\bf q_3}]&=&[(1, 0), (1, 1)]\cr
[{\bf q_4}]&=&[(2, 1)]
\eea
and the matrices $C$ and $\tilde{C}$ are
\bea
C=\left[\begin{array}{cccc}
1 &1 &1 &1\\
0 &1 &0 &1\\
0 &0 &1 &2\\
0 &0 &0 &1
\end{array}\right]\qquad
\tilde{C}=\left[\begin{array}{cccc}
1 &-1 &-1 &2\\
0 &1  &0  &-1\\
0 &0  &1  &-2\\
0 &0  &0  &1
\end{array}\right]
\eea

Using the second $M_{[\bq]}$ basis the finite $K$ cutoff is easily appreciated.
The key idea is that as soon as we have more than $K$ sums, since $X$ and $Y$ are $K\times K$ matrices, there is
no way to avoid repeating indices in the sums and hence the corresponding $M_{[\bq]}$ vanishes. 
The linear combinations of traces which vanish at finite $K$ are obtained by setting to zero the $M_{[\bq]} $ corresponding 
to vector partitions $\bq$ with $M>K$ parts.  

The finite $K$ relations that we have described above are universal in the sense that they will be present for any group $G$.
It is also possible that there are additional relations that rely on specific properties of the group $G$ being considered.
A simple example to illustrates the point arises for the case of an Abelian group $G$. 
In this case every irreducible representation is one dimensional so that matrix $X$ in (\ref{explMats}) is
proportional to the identity matrix. 
This implies new relations including, for example
\bea
\Tr (X)^2=K\Tr (X^2)
\eea

We can introduce an inner product on the traces which is the projector for the $M_{[\bq]}$ within the bound. 
It is expressible in terms of the matrices $C\tilde C$.
We will leave a discussion of the general inner product on diagonal matrices compatible with the cut-off to the future.

\section{ 2D/3D holography and factorization puzzle.  }\label{sec:FNS3D}

Following the formulation of the stringy exclusion principle \cite{malstrom}, the integrality and finiteness of parameters, such as $N$ in $S^N(X)$ symmetric group orbifold CFTs 
or $N$ in $U(N)$ ${\cal N}=4$ SYM have played an important role in understanding aspects of the holographic duality map. In this paper, we have studied in detail the relations between amplitudes of $G$-TQFT2 which follow from the finiteness of the dimension of the centre (denoted $K$) of the group algebra of $G$. Since the closed string amplitudes for connected and disconnected surfaces are expressible in terms of powers of traces of a matrix $X$ of size $K$, there are universal relations depending on $K$ which follow from properties of multi-traces of finite matrices much as in $AdS_5/CFT_4$. The trace structures of multi-traces can be encoded in permutations:  these arise from permutations of matrix indices which result in the traces.  Permutation combinatorics  plays a central role in the mapping from gauge invariant operators to giant gravitons  in the half-BPS sector  \cite{CJR} and beyond \cite{KR1,BHR1,BHR2,BCD,BDS,Ryo,RobertGraphs,CLBSR,Heslop} (for a review see \cite{RamgoReview}).
 As explained in section \ref{sec:Kinnerproduct}
the finite $K$ relations in $G$-TQFT2 can be expressed as null states in inner products defined using  permutations. This naturally leads to a formulation of these inner products in terms of a $ \mC ( G ) \times ( \mC ( S) )_{ \infty } $ TQFT2 (section \ref{sec:KSinfty}). Here we discuss the possibility that this $ \mC ( G ) \times ( \mC ( S) )_{ \infty } $-TQFT2 has a 3D holographic dual and in this scenario consider the factorization puzzle around the interpretation of 2D/3D holography in the presence of wormholes \cite{MaldaMaoz}. The puzzle concerns 3D holographic quantum  gravitational theories which have a disconnected boundary consisting of multiple surfaces. If there is an AdS/CFT set-up, the expectation is that the CFT partition function factorizes while from the bulk it is expected that the existence of a common bulk leads to a non-factorizing partition function. Recent discussions of the puzzle include \cite{SSS,BeBo,AltSon,SSSY-2021-03}. Here we give a different perspective on the puzzle based on the constructions in this paper which does not rely on ensembles or randomness but rather on the distinction between different types of observables within a hypothetical holographic dual of the constructions given earlier in the paper.

We consider the scenario where the  $ \mC( G ) \times  (\mC ( S ) )_{ \infty } $-TQFT2 theory, as used in the finiteness Section \ref{sec:finiteK}, has a 3D holographic dual. The theory contains probabilities for multiple circles going to multiple cirlces as in section \ref{sec:probabilities} : the transitions can proceed via fixed genus surfaces in $G$-TQFT2.  Cutting the surfaces to excise  a disc and inserting projectors $P_R$ calculates the probabilities for the $R$-sectors which weight the boundary conditions on the circles. 
By considering observables involving  the $(\mC ( S ) )_{ \infty }  $ sector (see Figure \ref{fig:fullinnerproduct}), we construct the inner products between in- and out- closed surface states. These inner products have a factorization property at large $K$
but there are $1/K$ corrections which cause mixing between surfaces. This is a $G$-TQFT2 analog of the failure of 
large $N$ factorization of traces which was observed to have important implications for the AdS/CFT map for large operators \cite{BBNS}. 

The scenario of  $ \mC( G ) \times  (\mC ( S ) )_{ \infty } $-TQFT2 having a holographic dual which includes wormholes is one where ensembles are not necessary to accommodate the existence of wormholes, but rather different choices of observables  within a single quantum theory lead to amplitudes which factorize or not. Observables involving just the $T_p$ observables associated to conjugacy classes $ \cC_p$ (as used in section \ref{sec:probabilities}) do factorize, since the $T_p$ observables can be inserted on disjoint surfaces and the independent boundary partition functions computed for example using the lattice formulation of $G$-TQFT2. By realising $G$ as monodromy groups of  covering spaces as in \cite{Mednykh} the $T_p$ observables can be interpreted in terms of winding string sectors along the lines of  \cite{GrTa}.  Observables involving the handle creation operators  $\Pi$ in $\cZ( \mC ( G ) ) $  \eqref{handlecreation} which typically involve many different conjugacy classes, coupled to the $ (\mC ( S ) )_{ \infty } $ sector  ( as in section \ref{sec:KSinfty})
 capture transition amplitudes between surfaces, which factorize at large $K$ but have $1/K$ corrections to factorization.  The distinction between the $T_p$ observables associated with fixed conjugacy classes and the observables such as the projectors $P_R$ which are sums over all conjugacy classes weighted by characters has played an important role in $AdS_5/CFT_4$ where the $T_p$ for symmetric groups can be associated to perturbative graviton states or low order multipole moments of the gravity field  \cite{IILoss}  while the $P_R$ can be associated to giant gravitons \cite{CJR}. This is used in \cite{IILoss} to formulate a model of information loss in the simplified set-up  of half-BPS states of ${\cal N}=4$ SYM and their gravitational duals \cite{LLM}.  
The ability of the $T_p$ to distinguish the different $P_R$ has led to a detailed study of the centre of the symmetric group algebra  \cite{KR1911} and associated Hamiltonians play a role in constructing Kronecker coefficients using ribbon graphs \cite{QMRibb}. Here we are adding to the class of interesting large operators relevant to holographic discussions the operators $ \Pi$ which create handles in $G$-TQFT2 or $G$-CTST. In fact this raises the interesting question of the interpretation in terms of LLM geometries in AdS5/CFT4 of the $\Pi$ operators for symmetric groups. The characterisation of small operators accessible to effective field theory and larger operators that can for example create black holes or access long-time evolution  in black hole evaporation is important for  holographic discussions of the black hole information paradox (see e.g. \cite{PapaRaju}).

To summarise  we are addressing the puzzle \cite{MaldaMaoz}, subject to the assumption  that there is a  gravitational 3D holographic dual for TQFT2 based on $ \mC( G ) \times  (\mC ( S ) )_{ \infty } $ of the kind we have described, which accounts for universal finiteness relations of $G$-TQFT2 as null states in an inner product.  Assuming such a dual exists, it is plausible that the intricate map between observables and topological interpretations in  the $ \mC( G ) \times  (\mC ( S ) )_{ \infty } $, allowing both factorizing and non-factorizing amplitudes, would have an analog in the bulk.

$G$-TQFT2 on a surface $ \Sigma$ has close relations to 3D topological theory on $ \Sigma \times S^1$ based on lattice constructions for quantizing Chern Simons theory \cite{AGS1,AGS2,BR1995}. A general discussion of lattice topological field theory with Hopf algebras extending these works  has been given \cite{CMDW1512,CM1607}, which also encompasses the Kitaev model developed for applications in quantum computing \cite{Kitaev}.  An interesting question is whether these constructions can be used to gain insights on a possible  holographic dual of TQFT2 based on $ \mC( G ) \times  (\mC ( S ) )_{ \infty } $  and its implications for the factorization puzzle.

\section{  Summary   and outlook   }

In this paper, we have developed   links between group theoretic
computational algorithms for dimensions and characters of  finite groups $G$ and $G$-TQFT2. 
We observed, in particular,  that the integer ratios $(|G|/d_R)^2$, where $d_R$ is the dimension of irreducible rep $R$ can be constructed by combinatoric algorithms which take as input the amplitudes of $G$-TQFT2 on  surfaces. These ratios enter the expansion of the handle creation operator in the basis of projection operators for the centre of the group algebra, $\cZ(\mC (G))$. The relation between the projector basis and the conjugacy class basis of $ \cZ( \mC ( G))$ plays a key role in the algorithms as well as the geometrical gluing properties that define $G$-TQFT2. 
Summing the amplitudes of $G$-TQFT2 weighted by a string coupling defines   combinatoric topological  string theory \cite{MarMax,GardMeg}, which we call $G$-CTST. We studied $S$-duality and the analytic structure of $G$-CTST as a function of the string coupling, connecting these to group theoretic combinatoric data. 

The two-dimensional path integral interpretation of $G$-TQFT2, which is  evident in its topological lattice formulation and is also central to its understanding as an example realizing Atiyah's axioms of TQFT, leads to the definition of a number of probability distributions. We described these and their inter-relations. This discussion included the Plancherel distribution for finite groups in mathematics \cite{BorOk} and made contact, by regarding the 2D theory as a model for 2D wormholes along the lines of \cite{MarMax},  with  Coleman's $\alpha$-states of wormhole physics \cite{Coleman:1988cy}. We explained that the Hilbert space structure of  $\cZ(\mC (G)) =\cH$ and the associated tower of symmetrised tensor products 
\bea 
S^{ \infty } ( \cH) = \bigoplus_{n=0}^{ \infty } S^n ( \cH) 
\eea
can be viewed as a topological quantum mechanics underlying these probability distributions. 

We have encoded the finite $K$ relations in $G$-TQFT2/$G$-CTST using an inner product (one of a family of 
possible inner products) on a polynomial algebra of surfaces. The definition of this inner product draws on the fact that $G$-TQFT2 amplitudes can be expressed in terms of  traces of powers of matrices of size $K$, which is equal to the dimension of $\cZ(\mC (G))$. This allows us to exploit the mathematics of finite $N$ effects in AdS5/CFT4 involving $U(N)$ gauge theory, which inform the physics of giant gravitons.  These finite $N$ effects are encoded in inner products based on symmetric group combinatorics, which arises since permutations are used to contract indices of matrices to build $U(N)$ gauge invariants. 
This led to a geometrical 
interpretation of the inner product by coupling $G$-TQFT2 to TQFT for a tower of symmetric group algebras 
\bea
( \mC(S))_{ \infty} \equiv 
\bigoplus_{ n  =0}^{ \infty } \mC ( S_n  )   
\eea
We have considered the scenario where all the observables  of the coupled TQFT2 theory of involving $\mC( G ) \times 
( \mC(S))_{ \infty} $ have a 3D holographic dual. In this scenario we discussed the factorization puzzle of \cite{MaldaMaoz}. The coupled TQFT2 contains both factorising and non-factorising amplitudes for products of surfaces, depending on the choice of observables. Assuming it has a 3D   holographic dual, there would be both types of amplitudes in the bulk theory, with appropriate  (possibly subtle) choice of boundary conditions.

The physical and mathematical aspects of our discussion should admit generalizations based on enlarging the considerations from the centre $ \cZ( \mC (G))$ to the full group algebra $ \mC( G )$, and considering the full  open-closed theory \cite{MooreSegal}. 
 Using the  links to group-theoretic algorithms  of the kind we have developed (which each come with their computational complexity characteristics), it will be interesting to investigate to what extent the interplay between wormholes, computational complexity and hierarchies of Hilbert spaces which played a role in this paper  generalizes to higher dimensional wormhole physics. In particular, it will be very interesting to make contact with discussions of complexity and black holes \cite{CompBH}.  

The reconstruction of representation theoretic quantities from group multiplication data which formed the focus of the first part  of the paper is  closely related to the concept of representation theory as a tool for  non-Abelian Fourier transforms for groups. Combinatoric Toplogical String Theory ($G$-CTST) seems to be an interesting toolkit for  geometrical constructive algorithms realising Fourier transforms in group theory.  
It raises the question of whether physical string theories could be interpreted in an analogous manner as providing the data for the construction of appropriate transforms, yet to be described.   

\begin{center} 
{\bf Acknowledgements}
\end{center} 
SR is supported by the STFC consolidated grant ST/P000754/1 `` String Theory, Gauge Theory \& Duality” and  a Visiting Professorship at the University of the Witwatersrand, funded by a Simons Foundation grant (509116)  awarded to the Mandelstam Institute for Theoretical Physics.
RdMK is supported by the Science and Technology Program of Guangzhou (No. 2019050001 and
No. 2020A1515010388), by the National Natural Science Foundation of China under Grant No. 12022512 and No. 12035007,
by a Simons Foundation Grant Award ID 509116 and by the South African Research Chairs initiative 
of the Department of Science and Technology and the National Research Foundation.
 We thank George Barnes, Joseph Ben Geloun,
Adrian Padellaro and Eric Sharpe for useful discussions  on the subject of this paper.

\appendix

\section{Normalized Characters}

The algorithm outlined in section \ref{CombConsGdR} can also be used to construct normalized characters of $T_{p}$, where $p$ labels the conjugacy class $\mathcal{C}_{p}$ of group $G$. By knowing the quantities $\mathrm{tr}(X^{r}_{p})$, which are sums over the group irreps of powers of the normalized characters, the algorithm describes how to construct the individual normalized characters by looking for the intersection of the sets of divisors of the polynomials $F(X_{p}, x = 0 ), F(X_{p} , x = 1)$ and so on.
In this appendix we generate sequences for the normalized characters of $T_{p}$, where $p$ labels the conjugacy classes of the symmetric group $S_{n}$, summed over the irreps of $S_{n}$ for $n = 1, 2, 3, \cdots, 20$. Table \ref{Tab:NormCharacTable} shows the sums over normalized characters for $T_{(3)}$, $T_{(2,2)}$ $T_{(5)}$, and $T_{(n)}$.

\begin{table}[h!]
\begin{center}
\begin{tabular}{|c|c|c|c|c|}
\hline
 $n$ & $\sum_{R}\frac{\chi_{R}(T_{(3)})}{d_{R}}$ & $\sum_{R}\frac{\chi_{R}(T_{(2,2)})}{d_{R}}$ & $\sum_{R}\frac{\chi_{R}(T_{(5)})}{d_{R}}$ & $\sum_{R}\frac{\chi_{R}(T_{(n)})}{d_{R}}$ \\
 \hline
 1 & 0&0 & 0 & 0 \\
 \hline
 2 & 0&0 & 0 & 0 \\
 \hline
 3 & 3&0 & 0  & 3 \\
 \hline
 4 & 12&7 & 0 & 0 \\
 \hline
 5 & 42&31 & 40 &40 \\
 \hline
 6 & 99 &118 & 265 &  0 \\
 \hline
 7 & 231 & 309 & 1080 & 1260 \\
 \hline
 8 & 462 & 772 & 3270 & 0 \\
 \hline
 9 & 882 & 1642 & 8900 & 72576 \\
 \hline
 10 & 1596 & 3391 & 20600 & 0 \\
 \hline
 11 & 2772 & 6348 & 45360 & 6652800 \\
 \hline
 12 & 4620 & 11779 & 91440 & 0 \\
 \hline
 13 & 7524 & 20317 & 177540 & 889574400 \\
 \hline
 14 & 11949 & 34849 & 325475 & 0 \\
 \hline
 15 & 18480 & 56923 & 581380 & 163459296000 \\
 \hline
 16 & 28182 & 92314 & 997670 & 0 \\
 \hline
 17 & 42108 & 144178 & 1676000 &  39520825344000 \\
 \hline
 18 & 62139 & 224425 & 2733785 & 0 \\
 \hline
 19 & 90216 & 338611 & 4384100 & 12164510040883200 \\
 \hline
 20 & 129690 & 509153 & 6875830 & 0 \\
 \hline
 \end{tabular}
\end{center}
\caption{Table listing the sequences of sums of normalized characters of $T_{p}$, where $p$ label conjugacy classes of the symmetric group $S_{n}$. We consider the conjugacy classes $p = (3), (2,2), (5)$ and $(n)$ for $n=1$ up to $n = 20$. The final column shows the normalized character for $T_{(3)}$ for $n = 3$, the normalized character for $T_{(5)}$ for $n=5$ and so on. These quantities for even $n$ are all zero.}
\label{Tab:NormCharacTable}
\end{table}

\vfill
\eject

\section{Non-Abelian groups up to size 60}\label{sec:NAB60}
{\tiny
\begin{longtable}{c|c|>{$}l<{$}}
\text{Group} & \text{Size} & P \\ \hline
 \text{S3} & 6 & 1 \\
 \text{D8} & 8 & 1 \\
 \text{Q8} & 8 & 1 \\
 \text{D10} & 10 & x^2+x-1 \\
 \text{C3 : C4} & 12 & x^2+9 \\
 \text{A4} & 12 & x^2+4 x+16 \\
 \text{D12} & 12 & 1 \\
 \text{D14} & 14 & x^3+x^2-2 x-1 \\
 \text{(C4 x C2) : C2} & 16 & x^2+4 \\
 \text{C4 : C4} & 16 & x^2+4 \\
 \text{C8 : C2} & 16 & \left(x^2+1\right) \left(x^2+4\right) \\
 \text{D16} & 16 & x^2-2 \\
 \text{QD16} & 16 & x^2+2 \\
 \text{Q16} & 16 & x^2-2 \\
 \text{C2 x D8} & 16 & 1 \\
 \text{C2 x Q8} & 16 & 1 \\
 \text{(C4 x C2) : C2} & 16 & x^2+1 \\
 \text{D18} & 18 & x^3-3 x+1 \\
 \text{C3 x S3} & 18 & \left(x^2-3 x+9\right) \left(x^2-x+1\right) \left(x^2+x+1\right) \left(x^2+2 x+4\right) \left(x^2+3 x+9\right) \\
 \text{(C3 x C3) : C2} & 18 & 1 \\
 \text{C5 : C4} & 20 & \left(x^2+25\right) \left(x^2-x-1\right) \left(x^2+x-1\right) \\
 \text{C5 : C4} & 20 & x^2+25 \\
 \text{D20} & 20 & \left(x^2-x-1\right) \left(x^2+x-1\right) \\
 \text{C7 : C3} & 21 & \left(x^2+x+2\right) \left(x^2+7 x+49\right) \\
 \text{D22} & 22 & x^5+x^4-4 x^3-3 x^2+3 x+1 \\
 \text{C3 : C8} & 24 & \left(x^2+1\right) \left(x^2+4\right) \left(x^2+9\right) \left(x^4+81\right) \\
 \text{SL(2,3)} & 24 & \left(x^2-2 x+4\right) \left(x^2+2 x+4\right) \left(x^2+4 x+16\right) \\
 \text{C3 : Q8} & 24 & x^2-3 \\
 \text{C4 x S3} & 24 & \left(x^2+1\right) \left(x^2+4\right) \left(x^2+9\right) \\
 \text{D24} & 24 & x^2-3 \\
 \text{C2 x (C3 : C4)} & 24 & x^2+9 \\
 \text{(C6 x C2) : C2} & 24 & x^2+3 \\
 \text{C3 x D8} & 24 & \left(x^2-2 x+4\right) \left(x^2-x+1\right) \left(x^2+x+1\right) \left(x^2+2 x+4\right) \\
 \text{C3 x Q8} & 24 & \left(x^2-2 x+4\right) \left(x^2-x+1\right) \left(x^2+x+1\right) \left(x^2+2 x+4\right) \\
 \text{S4} & 24 & 1 \\
 \text{C2 x A4} & 24 & \left(x^2-4 x+16\right) \left(x^2+4 x+16\right) \\
 \text{C2 x C2 x S3} & 24 & 1 \\
 \text{D26} & 26 & x^6+x^5-5 x^4-4 x^3+6 x^2+3 x-1 \\
 \text{(C3 x C3) : C3} & 27 & \left(x^2+x+1\right) \left(x^2+3 x+9\right) \\
 \text{C9 : C3} & 27 & \left(x^2+x+1\right) \left(x^2+3 x+9\right) \\
 \text{C7 : C4} & 28 & \left(x^2+49\right) \left(x^3-x^2-2 x+1\right) \left(x^3+x^2-2 x-1\right) \\
 \text{D28} & 28 & \left(x^3-x^2-2 x+1\right) \left(x^3+x^2-2 x-1\right) \\
 \text{C5 x S3} & 30 &\left(x^4-3 x^3+9 x^2-27 x+81\right) \left(x^4-x^3+x^2-x+1\right)\left(x^4+x^3+x^2+x+1\right) \\
 & & \qquad\times\left(x^4+2 x^3+4 x^2+8 x+16\right) \left(x^4+3 x^3+9 x^2+27 x+81\right) \\
 \text{C3 x D10} & 30 & \left(x^2-5 x+25\right) \left(x^2+x-1\right) \left(x^2+x+1\right) \left(x^2+2 x+4\right) \left(x^2+5 x+25\right) \left(x^4-x^3+2 x^2+x+1\right) \\
 \text{D30} & 30 & \left(x^2+x-1\right) \left(x^4-x^3-4 x^2+4 x+1\right) \\
 \text{(C4 x C2) : C4} & 32 & x^2+4 \\
 \text{C8 : C4} & 32 & \left(x^2+1\right) \left(x^2+4\right) \\
 \text{(C8 x C2) : C2} & 32 & \left(x^2+1\right) \left(x^2+4\right) \left(x^4+16\right) \\
 \text{(C2 x C2 x C2) : C4} & 32 & x^2+16 \\
 \text{(C8 : C2) : C2} & 32 & x^2+16 \\
 \text{C2 . ((C4 x C2) : C2)} 
 	& 32 & x^2+16 \\
 \text{(C8 x C2) : C2} & 32 & \left(x^2-2\right) \left(x^2+2\right) \left(x^2+4\right) \left(x^2+16\right) \\
 \text{Q8 : C4} & 32 & \left(x^2-2\right) \left(x^2+2\right) \left(x^2+4\right) \left(x^2+16\right) \\
 \text{(C4 x C4) : C2} & 32 & \left(x^2+1\right) \left(x^2+4\right) \left(x^2+16\right) \left(x^2-2 x+2\right) \left(x^2+2 x+2\right) \\
 \text{C4 : C8} & 32 & \left(x^2+1\right) \left(x^2+4\right) \left(x^4+16\right) \\
 \text{C8 : C4} & 32 & \left(x^2+2\right) \left(x^2+16\right) \\
 \text{C8 : C4} & 32 & \left(x^2-2\right) \left(x^2+16\right) \\
 \text{C4 . D8 = C4 . (C4 x C2)} & 32 & \left(x^2-2\right) \left(x^2+1\right) \left(x^2+2\right) \left(x^2+16\right) \\
 \text{C16 : C2} & 32 & \left(x^2+1\right) \left(x^2+4\right) \left(x^4+1\right) \left(x^4+16\right) \\
 \text{D32} & 32 & \left(x^2-2\right) \left(x^4-4 x^2+2\right) \\
 \text{QD32} & 32 & \left(x^2-2\right) \left(x^4+4 x^2+2\right) \\
 \text{Q32} & 32 & \left(x^2-2\right) \left(x^4-4 x^2+2\right) \\
 \text{C2 x ((C4 x C2) : C2)} & 32 & x^2+4 \\
 \text{C2 x (C4 : C4)} & 32 & x^2+4 \\
 \text{(C4 x C4) : C2} & 32 & \left(x^2+1\right) \left(x^2+4\right) \\
 \text{C4 x D8} & 32 & \left(x^2+1\right) \left(x^2+4\right) \\
 \text{C4 x Q8} & 32 & \left(x^2+1\right) \left(x^2+4\right) \\
 \text{(C2 x C2 x C2 x C2) : C2} & 32 & 1 \\
 \text{(C4 x C2 x C2) : C2} & 32 & x^2+4 \\
 \text{(C2 x Q8) : C2} & 32 & x^2+4 \\
 \text{(C4 x C2 x C2) : C2} & 32 & x^2+4 \\
 \text{(C4 x C4) : C2} & 32 & x^2+4 \\
 \text{(C2 x C2) . (C2 x C2 x C2)} & 32 & x^2+4 \\
 \text{(C4 x C4) : C2} & 32 & x^2+4 \\
 \text{(C4 x C4) : C2} & 32 & 1 \\
 \text{C4 : Q8} & 32 & 1 \\
 \text{C2 x (C8 : C2)} & 32 & \left(x^2+1\right) \left(x^2+4\right) \\
 \text{(C8 x C2) : C2} & 32 & \left(x^2+1\right) \left(x^2+4\right) \left(x^4+1\right) \\
 \text{C2 x D16} & 32 & x^2-2 \\
 \text{C2 x QD16} & 32 & x^2+2 \\
 \text{C2 x Q16} & 32 & x^2-2 \\
 \text{(C8 x C2) : C2} & 32 & \left(x^2-2\right) \left(x^2+1\right) \left(x^2+2\right) \\
 \text{C8 : (C2 x C2)} & 32 & 1 \\
 \text{(C2 x Q8) : C2} & 32 & 1 \\
 \text{C2 x C2 x D8} & 32 & 1 \\
 \text{C2 x C2 x Q8} & 32 & 1 \\
 \text{C2 x ((C4 x C2) : C2)} & 32 & x^2+1 \\
 \text{(C2 x C2 x C2) : (C2 x C2)} & 32 & 1 \\
 \text{(C2 x Q8) : C2} & 32 & 1 \\
 \text{D34} & 34 & x^8+x^7-7 x^6-6 x^5+15 x^4+10 x^3-10 x^2-4 x+1 \\
 \text{C9 : C4} & 36 & \left(x^2+81\right) \left(x^3-3 x-1\right) \left(x^3-3 x+1\right) \\
 \text{(C2 x C2) : C9} & 36 & \left(x^2-x+1\right) \left(x^2+x+1\right) \left(x^2+3 x+9\right) \left(x^2+4 x+16\right) \left(x^6+64 x^3+4096\right) \\
 \text{D36} & 36 & \left(x^3-3 x-1\right) \left(x^3-3 x+1\right) \\
 \text{C3 x (C3 : C4)} & 36 & \left(x^2+9\right) \left(x^2-3 x+9\right) \left(x^2-2 x+4\right) \left(x^2-x+1\right) \left(x^2+x+1\right) \left(x^2+2 x+4\right) \\
& &\qquad\times \left(x^2+3 x+9\right)\left(x^4-9 x^2+81\right) \\
 \text{(C3 x C3) : C4} & 36 & x^2+81 \\
 \text{(C3 x C3) : C4} & 36 & x^2+81 \\
 \text{S3 x S3} & 36 & 1 \\
 \text{C3 x A4} & 36 & \left(x^2-x+1\right) \left(x^2+x+1\right) \left(x^2+3 x+9\right) \left(x^2+4 x+16\right) \\
 \text{C6 x S3} & 36 & \left(x^2-3 x+9\right) \left(x^2-2 x+4\right) \left(x^2-x+1\right) \left(x^2+x+1\right) \left(x^2+2 x+4\right) \left(x^2+3 x+9\right) \\
 \text{C2 x ((C3 x C3) : C2)} & 36 & 1 \\
 \text{D38} & 38 & x^9+x^8-8 x^7-7 x^6+21 x^5+15 x^4-20 x^3-10 x^2+5 x+1 \\
 \text{C13 : C3} & 39 & \left(x^2+13 x+169\right) \left(x^4+x^3+2 x^2-4 x+3\right) \\
 \text{C5 : C8} & 40 & \left(x^2+1\right) \left(x^2+4\right) \left(x^2+25\right) \left(x^2-x-1\right) \left(x^2+x-1\right) \left(x^4+625\right) \left(x^4+3 x^2+1\right) \\
 \text{C5 : C8} & 40 & \left(x^2+25\right) \left(x^4+625\right) \\
 \text{C5 : Q8} & 40 & \left(x^2-x-1\right) \left(x^2+x-1\right) \left(x^4-5 x^2+5\right) \\
 \text{C4 x D10} & 40 & \left(x^2+1\right) \left(x^2+4\right) \left(x^2+25\right) \left(x^2-x-1\right) \left(x^2+x-1\right) \left(x^4+3 x^2+1\right) \\
 \text{D40} & 40 & \left(x^2-x-1\right) \left(x^2+x-1\right) \left(x^4-5 x^2+5\right) \\
 \text{C2 x (C5 : C4)} & 40 & \left(x^2+25\right) \left(x^2-x-1\right) \left(x^2+x-1\right) \\
 \text{(C10 x C2) : C2} & 40 & \left(x^2-x-1\right) \left(x^2+x-1\right) \left(x^4+5 x^2+5\right) \\
 \text{C5 x D8} & 40 & \left(x^4-2 x^3+4 x^2-8 x+16\right) \left(x^4-x^3+x^2-x+1\right) \left(x^4+x^3+x^2+x+1\right)\\ & &\qquad\times \left(x^4+2 x^3+4 x^2+8 x+16\right) \\
 \text{C5 x Q8} & 40 & \left(x^4-2 x^3+4 x^2-8 x+16\right) \left(x^4-x^3+x^2-x+1\right) \left(x^4+x^3+x^2+x+1\right)\\ & &\qquad\times \left(x^4+2 x^3+4 x^2+8 x+16\right) \\
 \text{C2 x (C5 : C4)} & 40 & x^2+25 \\
 \text{C2 x C2 x D10} & 40 & \left(x^2-x-1\right) \left(x^2+x-1\right) \\
 \text{C7 : C6} & 42 & \left(x^2-7 x+49\right) \left(x^2+7 x+49\right) \\
 \text{C2 x (C7 : C3)} & 42 & \left(x^2-7 x+49\right) \left(x^2-x+2\right) \left(x^2+x+2\right) \left(x^2+7 x+49\right) \\
 \text{C7 x S3} & 42 & \left(x^6-3 x^5+9 x^4-27 x^3+81 x^2-243 x+729\right) \left(x^6-x^5+x^4-x^3+x^2-x+1\right)\\
& &\qquad\times  \left(x^6+x^5+x^4+x^3+x^2+x+1\right)\left(x^6+2 x^5+4 x^4+8 x^3+16 x^2+32 x+64\right)\\
& &\qquad\times \left(x^6+3 x^5+9 x^4+27 x^3+81 x^2+243 x+729\right) \\
 \text{C3 x D14} & 42 & \left(x^2-7 x+49\right) \left(x^2+x+1\right) \left(x^2+2 x+4\right) \left(x^2+7 x+49\right) \left(x^3+x^2-2 x-1\right)\\
& &\qquad\times \left(x^6-x^5+3 x^4+5 x^2-2 x+1\right) \\
 \text{D42} & 42 & \left(x^3+x^2-2 x-1\right) \left(x^6-x^5-6 x^4+6 x^3+8 x^2-8 x+1\right) \\
 \text{C11 : C4} & 44 & \left(x^2+121\right) \left(x^5-x^4-4 x^3+3 x^2+3 x-1\right) \left(x^5+x^4-4 x^3-3 x^2+3 x+1\right) \\
 \text{D44} & 44 & \left(x^5-x^4-4 x^3+3 x^2+3 x-1\right) \left(x^5+x^4-4 x^3-3 x^2+3 x+1\right) \\
 \text{D46} & 46 & x^{11}+x^{10}-10 x^9-9 x^8+36 x^7+28 x^6-56 x^5-35 x^4+35 x^3+15 x^2-6 x-1 \\
 \text{C3 : C16} & 48 & \left(x^2+1\right) \left(x^2+4\right) \left(x^2+9\right) \left(x^4+1\right) \left(x^4+16\right) \left(x^4+81\right) \left(x^8+6561\right) \\
 \text{(C4 x C4) : C3} & 48 & \left(x^2+2 x+5\right) \left(x^2+16 x+256\right) \\
 \text{C8 x S3} & 48 & \left(x^2+1\right) \left(x^2+4\right) \left(x^2+9\right) \left(x^4+1\right) \left(x^4+16\right) \left(x^4+81\right) \\
 \text{C24 : C2} & 48 & \left(x^2+1\right) \left(x^2+4\right) \left(x^2+36\right) \left(x^4+9\right) \\
 \text{C24 : C2} & 48 & \left(x^2-3\right) \left(x^2+2\right) \left(x^4+4 x^2+1\right) \\
 \text{D48} & 48 & \left(x^2-3\right) \left(x^2-2\right) \left(x^4-4 x^2+1\right) \\
 \text{C3 : Q16} & 48 & \left(x^2-3\right) \left(x^2-2\right) \left(x^4-4 x^2+1\right) \\
 \text{C2 x (C3 : C8)} & 48 & \left(x^2+1\right) \left(x^2+4\right) \left(x^2+9\right) \left(x^4+81\right) \\
 \text{(C3 : C8) : C2} & 48 & \left(x^2-3\right) \left(x^2+1\right) \left(x^2+3\right) \left(x^2+4\right) \left(x^2+36\right) \\
 \text{C4 x (C3 : C4)} & 48 & \left(x^2+1\right) \left(x^2+4\right) \left(x^2+9\right) \\
 \text{(C3 : C4) : C4} & 48 & \left(x^2-3\right) \left(x^2+1\right) \left(x^2+3\right) \left(x^2+4\right) \left(x^2+36\right) \\
 \text{C12 : C4} & 48 & \left(x^2-3\right) \left(x^2+36\right) \\
 \text{(C12 x C2) : C2} & 48 & \left(x^2-3\right) \left(x^2+1\right) \left(x^2+3\right) \left(x^2+4\right) \left(x^2+36\right) \\
 \text{(C3 x D8) : C2} & 48 & \left(x^2-18\right) \left(x^2+12\right) \\
 \text{(C3 : Q8) : C2} & 48 & \left(x^2+12\right) \left(x^2+18\right) \\
 \text{(C3 x Q8) : C2} & 48 & \left(x^2+12\right) \left(x^2+18\right) \\
 \text{C3 : Q16} & 48 & \left(x^2-18\right) \left(x^2+12\right) \\
 \text{(C6 x C2) : C4} & 48 & \left(x^2+3\right) \left(x^2+36\right) \\
 \text{C3 x ((C4 x C2) : C2)} & 48 & \left(x^2+4\right) \left(x^2-2 x+4\right) \left(x^2-x+1\right) \left(x^2+x+1\right) \left(x^2+2 x+4\right) \left(x^4-4 x^2+16\right) \\
 \text{C3 x (C4 : C4)} & 48 & \left(x^2+4\right) \left(x^2-2 x+4\right) \left(x^2-x+1\right) \left(x^2+x+1\right) \left(x^2+2 x+4\right) \left(x^4-4 x^2+16\right) \\
 \text{C3 x (C8 : C2)} & 48 & \left(x^2+1\right) \left(x^2+4\right) \left(x^2-2 x+4\right) \left(x^2-x+1\right) \left(x^2+x+1\right) \left(x^2+2 x+4\right) \left(x^4-4 x^2+16\right)\\
& &\qquad\times \left(x^4-x^2+1\right) \\
 \text{C3 x D16} & 48 & \left(x^2-2\right) \left(x^2-4 x+16\right) \left(x^2-2 x+4\right) \left(x^2-x+1\right) \left(x^2+x+1\right) \left(x^2+2 x+4\right) \\
& &\qquad\times \left(x^2+4 x+16\right)\left(x^4+2 x^2+4\right) \\
 \text{C3 x QD16} & 48 & \left(x^2+2\right) \left(x^2-4 x+16\right) \left(x^2-2 x+4\right) \left(x^2-x+1\right) \left(x^2+x+1\right) \left(x^2+2 x+4\right) \\
& &\qquad\times \left(x^2+4 x+16\right)\left(x^4-2 x^2+4\right) \\
 \text{C3 x Q16} & 48 & \left(x^2-2\right) \left(x^2-4 x+16\right) \left(x^2-2 x+4\right) \left(x^2-x+1\right) \left(x^2+x+1\right) \left(x^2+2 x+4\right) \\
& &\qquad\times \left(x^2+4 x+16\right)\left(x^4+2 x^2+4\right) \\
 \text{C2 . S4 = SL(2,3) . C2} & 48 & x^2-18 \\
 \text{GL(2,3)} & 48 & x^2+18 \\
 \text{A4 : C4} & 48 & \left(x^2+4\right) \left(x^2+36\right) \\
 \text{C4 x A4} & 48 & \left(x^2+1\right) \left(x^2+9\right) \left(x^2+16\right) \left(x^2-4 x+16\right) \left(x^2+4 x+16\right) \left(x^4-16 x^2+256\right) \\
 \text{C2 x SL(2,3)} & 48 & \left(x^2-4 x+16\right) \left(x^2-2 x+4\right) \left(x^2+2 x+4\right) \left(x^2+4 x+16\right) \\
 \text{((C4 x C2) : C2) : C3} & 48 & \left(x^2+1\right) \left(x^2+4\right) \left(x^2-4 x+16\right) \left(x^2-2 x+4\right) \left(x^2+2 x+4\right) \left(x^2+4 x+16\right) \left(x^4-4 x^2+16\right) \\
 \text{C2 x (C3 : Q8)} & 48 & x^2-3 \\
 \text{C2 x C4 x S3} & 48 & \left(x^2+1\right) \left(x^2+4\right) \left(x^2+9\right) \\
 \text{C2 x D24} & 48 & x^2-3 \\
 \text{(C12 x C2) : C2} & 48 & \left(x^2-3\right) \left(x^2+1\right) \left(x^2+3\right) \left(x^2+4\right) \\
 \text{D8 x S3} & 48 & 1 \\
 \text{(C4 x S3) : C2} & 48 & x^2+9 \\
 \text{Q8 x S3} & 48 & 1 \\
 \text{(C4 x S3) : C2} & 48 & x^2+9 \\
 \text{C2 x C2 x (C3 : C4)} & 48 & x^2+9 \\
 \text{C2 x ((C6 x C2) : C2)} & 48 & x^2+3 \\
 \text{C6 x D8} & 48 & \left(x^2-2 x+4\right) \left(x^2-x+1\right) \left(x^2+x+1\right) \left(x^2+2 x+4\right) \\
 \text{C6 x Q8} & 48 & \left(x^2-2 x+4\right) \left(x^2-x+1\right) \left(x^2+x+1\right) \left(x^2+2 x+4\right) \\
 \text{C3 x ((C4 x C2) : C2)} & 48 & \left(x^2+1\right) \left(x^2-2 x+4\right) \left(x^2-x+1\right) \left(x^2+x+1\right) \left(x^2+2 x+4\right) \left(x^4-x^2+1\right) \\
 \text{C2 x S4} & 48 & 1 \\
 \text{C2 x C2 x A4} & 48 & \left(x^2-4 x+16\right) \left(x^2+4 x+16\right) \\
 \text{(C2 x C2 x C2 x C2) : C3} & 48 & x^2+16 x+256 \\
 \text{C2 x C2 x C2 x S3} & 48 & 1 \\
 \text{D50} & 50 & \left(x^2+x-1\right) \left(x^{10}-10 x^8+35 x^6+x^5-50 x^4-5 x^3+25 x^2+5 x-1\right) \\
 \text{C5 x D10} & 50 & \left(x^2+x-1\right) \left(x^4-5 x^3+25 x^2-125 x+625\right) \left(x^4-3 x^3+4 x^2-2 x+1\right) \left(x^4+x^3+x^2+x+1\right)\\
& &\qquad\times \left(x^4+2 x^3+4 x^2+3 x+1\right) \left(x^4+2 x^3+4 x^2+8 x+16\right) \left(x^4+5 x^3+25 x^2+125 x+625\right) \\
 \text{(C5 x C5) : C2} & 50 & x^2+x-1 \\
 \text{C13 : C4} & 52 & \left(x^2+169\right) \left(x^6-x^5-5 x^4+4 x^3+6 x^2-3 x-1\right) \left(x^6+x^5-5 x^4-4 x^3+6 x^2+3 x-1\right) \\
 \text{C13 : C4} & 52 & \left(x^2+169\right) \left(x^3+x^2-4 x+1\right) \\
 \text{D52} & 52 & \left(x^6-x^5-5 x^4+4 x^3+6 x^2-3 x-1\right) \left(x^6+x^5-5 x^4-4 x^3+6 x^2+3 x-1\right) \\
 \text{D54} & 54 & \left(x^3-3 x+1\right) \left(x^9-9 x^7+27 x^5-30 x^3+9 x+1\right) \\
 \text{C3 x D18} & 54 & \left(x^2-9 x+81\right) \left(x^2-x+1\right) \left(x^2+x+1\right) \left(x^2+2 x+4\right) \left(x^2+9 x+81\right) \left(x^3-3 x+1\right)\\
& &\qquad\times \left(x^6+3 x^4+2 x^3+9 x^2+3 x+1\right) \\
 \text{C9 x S3} & 54 & \left(x^2-3 x+9\right) \left(x^2-x+1\right) \left(x^2+x+1\right) \left(x^2+2 x+4\right) \left(x^2+3 x+9\right) \left(x^6-27 x^3+729\right) \\
& & \qquad\times \left(x^6-x^3+1\right)\left(x^6+x^3+1\right) \left(x^6+8 x^3+64\right) \left(x^6+27 x^3+729\right)\\
 \text{(C3 x C3) : C6} & 54 & \left(x^2-9 x+81\right) \left(x^2-3 x+9\right) \left(x^2+3 x+9\right) \left(x^2+6 x+36\right) \left(x^2+9 x+81\right) \\
 \text{C9 : C6} & 54 & \left(x^2-9 x+81\right) \left(x^2-3 x+9\right) \left(x^2+3 x+9\right) \left(x^2+6 x+36\right) \left(x^2+9 x+81\right) \\
 \text{(C9 x C3) : C2} & 54 & x^3-3 x+1 \\
 \text{((C3 x C3) : C3) : C2} & 54 & \left(x^2-3 x+9\right) \left(x^2+x+1\right) \left(x^2+3 x+9\right) \\
 \text{C2 x ((C3 x C3) : C3)} & 54 & \left(x^2-3 x+9\right) \left(x^2-x+1\right) \left(x^2+x+1\right) \left(x^2+3 x+9\right) \\
 \text{C2 x (C9 : C3)} & 54 & \left(x^2-3 x+9\right) \left(x^2-x+1\right) \left(x^2+x+1\right) \left(x^2+3 x+9\right) \\
 \text{C3 x C3 x S3} & 54 & \left(x^2-3 x+9\right) \left(x^2-x+1\right) \left(x^2+x+1\right) \left(x^2+2 x+4\right) \left(x^2+3 x+9\right) \\
 \text{C3 x ((C3 x C3) : C2)} & 54 & \left(x^2-9 x+81\right) \left(x^2-x+1\right) \left(x^2+x+1\right) \left(x^2+2 x+4\right) \left(x^2+9 x+81\right) \\
 \text{(C3 x C3 x C3) : C2} & 54 & 1 \\
 \text{C11 : C5} & 55 & \left(x^2+x+3\right) \left(x^4+11 x^3+121 x^2+1331 x+14641\right) \\
 \text{C7 : C8} & 56 & \left(x^2+1\right) \left(x^2+4\right) \left(x^2+49\right) \left(x^3-x^2-2 x+1\right) \left(x^3+x^2-2 x-1\right) \left(x^4+2401\right) \left(x^6+5 x^4+6 x^2+1\right) \\
 \text{C7 : Q8} & 56 & \left(x^3-x^2-2 x+1\right) \left(x^3+x^2-2 x-1\right) \left(x^6-7 x^4+14 x^2-7\right) \\
 \text{C4 x D14} & 56 & \left(x^2+1\right) \left(x^2+4\right) \left(x^2+49\right) \left(x^3-x^2-2 x+1\right) \left(x^3+x^2-2 x-1\right) \left(x^6+5 x^4+6 x^2+1\right) \\
 \text{D56} & 56 & \left(x^3-x^2-2 x+1\right) \left(x^3+x^2-2 x-1\right) \left(x^6-7 x^4+14 x^2-7\right) \\
 \text{C2 x (C7 : C4)} & 56 & \left(x^2+49\right) \left(x^3-x^2-2 x+1\right) \left(x^3+x^2-2 x-1\right) \\
 \text{(C14 x C2) : C2} & 56 & \left(x^3-x^2-2 x+1\right) \left(x^3+x^2-2 x-1\right) \left(x^6+7 x^4+14 x^2+7\right) \\
 \text{C7 x D8} & 56 & \left(x^6-2 x^5+4 x^4-8 x^3+16 x^2-32 x+64\right) \left(x^6-x^5+x^4-x^3+x^2-x+1\right)\\ 
& &\qquad\times \left(x^6+x^5+x^4+x^3+x^2+x+1\right) \left(x^6+2 x^5+4 x^4+8 x^3+16 x^2+32 x+64\right) \\
 \text{C7 x Q8} & 56 & \left(x^6-2 x^5+4 x^4-8 x^3+16 x^2-32 x+64\right) \left(x^6-x^5+x^4-x^3+x^2-x+1\right)\\
& &\qquad\times \left(x^6+x^5+x^4+x^3+x^2+x+1\right)\left(x^6+2 x^5+4 x^4+8 x^3+16 x^2+32 x+64\right) \\
 \text{(C2 x C2 x C2) : C7} & 56 & x^6+8 x^5+64 x^4+512 x^3+4096 x^2+32768 x+262144 \\
 \text{C2 x C2 x D14} & 56 & \left(x^3-x^2-2 x+1\right) \left(x^3+x^2-2 x-1\right) \\
 \text{C19 : C3} & 57 & \left(x^2+19 x+361\right) \left(x^6+x^5+2 x^4-8 x^3-x^2+5 x+7\right) \\
 \text{D58} & 58 & x^{14}+x^{13}-13 x^{12}-12 x^{11}+66 x^{10}+55 x^9-165 x^8-120 x^7+210 x^6+126 x^5-126 x^4\\
& &\qquad -56 x^3+28 x^2+7 x-1 \\
 \text{C5 x (C3 : C4)} & 60 & \left(x^2+9\right) \left(x^4-3 x^3+9 x^2-27 x+81\right) \left(x^4-2 x^3+4 x^2-8 x+16\right) \left(x^4-x^3+x^2-x+1\right)\\
& &\qquad\times \left(x^4+x^3+x^2+x+1\right) \left(x^4+2 x^3+4 x^2+8 x+16\right) \left(x^4+3 x^3+9 x^2+27 x+81\right)\\
& &\qquad\times \left(x^8-9 x^6+81 x^4-729 x^2+6561\right) \\
 \text{C3 x (C5 : C4)} & 60 & \left(x^2+25\right) \left(x^2-5 x+25\right) \left(x^2-2 x+4\right) \left(x^2-x-1\right) \left(x^2-x+1\right) \left(x^2+x-1\right) \left(x^2+x+1\right)\\
& &\qquad\times \left(x^2+2 x+4\right) \left(x^2+5 x+25\right) \left(x^4-25 x^2+625\right) 
\left(x^4-x^3+2 x^2+x+1\right)\\
& &\qquad\times \left(x^4+x^3+2 x^2-x+1\right) \\
 \text{C15 : C4} & 60 & \left(x^2+225\right) \left(x^2-x-1\right) \left(x^2+x-1\right) \left(x^4-x^3-4 x^2+4 x+1\right) \left(x^4+x^3-4 x^2-4 x+1\right) \\
 \text{A5} & 60 & x^2-4 x-16 \\
 \text{C3 x (C5 : C4)} & 60 & \left(x^2+25\right) \left(x^2-5 x+25\right) \left(x^2-x+1\right) \left(x^2+x+1\right) \left(x^2+4 x+16\right) \left(x^2+5 x+25\right) \left(x^4-25 x^2+625\right) \\
 \text{C15 : C4} & 60 & \left(x^2+225\right) \left(x^2-x+4\right) \\
 \text{S3 x D10} & 60 & \left(x^2-3 x-9\right) \left(x^2-x-1\right) \left(x^2+x-1\right) \left(x^2+2 x-4\right) \left(x^2+3 x-9\right) \\
 \text{C5 x A4} & 60 & \left(x^2+4 x+16\right) \left(x^4-x^3+x^2-x+1\right) \left(x^4+x^3+x^2+x+1\right) \left(x^4+3 x^3+9 x^2+27 x+81\right)\\
& &\qquad\times \left(x^4+4 x^3+16 x^2+64 x+256\right) \left(x^8-4 x^7+64 x^5-256 x^4+1024 x^3-16384 x+65536\right) \\
 \text{C6 x D10} & 60 & \left(x^2-5 x+25\right) \left(x^2-2 x+4\right) \left(x^2-x-1\right) \left(x^2-x+1\right) \left(x^2+x-1\right) \left(x^2+x+1\right) \left(x^2+2 x+4\right)\\
& &\qquad\times \left(x^2+5 x+25\right) \left(x^4-x^3+2 x^2+x+1\right) \left(x^4+x^3+2 x^2-x+1\right) \\
 \text{C10 x S3} & 60 & \left(x^4-3 x^3+9 x^2-27 x+81\right) \left(x^4-2 x^3+4 x^2-8 x+16\right) \left(x^4-x^3+x^2-x+1\right) \\
& &\qquad\times 
\left(x^4+x^3+x^2+x+1\right) \left(x^4+2 x^3+4 x^2+8 x+16\right) \left(x^4+3 x^3+9 x^2+27 x+81\right) \\
 \text{D60} & 60 & \left(x^2-x-1\right) \left(x^2+x-1\right) \left(x^4-x^3-4 x^2+4 x+1\right) \left(x^4+x^3-4 x^2-4 x+1\right) \\
\end{longtable}
}

\section{Mathieu Groups}
Having presented the polynomials for first 60 non-Abelian groups, it is illustrative to attempt the other extreme of very large, non-trivial groups. Naturally, the Sporadics come to mind; these have become very much studied recently in the context of partition functions in quantum field theories.
Whilst the sizes of these groups could become astronomical, the number of conjugacy classes is very tame. For instance, the Monster, with size $\sim 10^{54}$, has only 194 classes.

Thus, extracting our polynomials for all 26 Sporadics is relatively easy.
Of course, the one for the Monster is still a bit too long to present.
Nevertheless, for the two most famous Sporadics in physics, viz, Mathieu 24 and 23, we have
\begin{equation}
\begin{array}{rcl}
	P_{M_{24}} & = &
	\left(x^2-388608 x+302032355328\right) \left(x^2-11264 x+253755392\right) \left(x^2+5888 x+69337088\right)\\ &&\left(x^2+11264 x+253755392\right) \left(x^2+11776 x+277348352\right) \left(x^2+13824 x+1146617856\right)\\ &&\left(x^2+17664 x+624033792\right) \left(x^2+70656 x+19969081344\right) \left(x^2+129536 x+33559150592\right) \\
	&&\left(x^2+259072 x+134236602368\right)\\
	P_{M_{23}} & = &
	\left(x^2-16192 x+524361728\right) \left(x^2+576 x+1990656\right) \left(x^2+736 x+1083392\right)\\ &&\left(x^2+1035 x+3213675\right) \left(x^2+2944 x+34668544\right) \left(x^2+16192 x+524361728\right)
	\ .
	\end{array}
\end{equation}

\end{document}